\documentclass[superscriptaddress,10pt,twocolumn,floatfix]{revtex4-2} 

\usepackage{mathtools}

\usepackage{hyperref}

\usepackage{array}[=2016-10-06]
\usepackage{booktabs,makecell}
\newcolumntype{P}[1]{>{\raggedright\arraybackslash}p{#1}}

\usepackage{amsmath}  
\usepackage{amsfonts} 
\usepackage{graphicx} 
\usepackage{esvect}
\usepackage{mathtools}

\usepackage{xcolor}
\usepackage{appendix}
\usepackage{siunitx} 
\usepackage{mathrsfs}

\usepackage{placeins}

\newcounter{researchquestion}

\usepackage[caption=false]{subfig} 

\usepackage{pbox}

\usepackage{mathtools}
\makeatletter
\newcommand{\vast}{\bBigg@{3}}
\newcommand{\Vast}{\bBigg@{3.5}}
\makeatother

\usepackage{etoolbox}

\makeatletter
\patchcmd{\@author@join@}
  {\@join{\@separator}{#2}{#3}}
  {#2\@if@empty{#2}{}{\textsuperscript{,}}%
   \frontmatter@footnote{#3}}
  {}{\PackageError{authornotes}{Author-note patch failed}{}}

\patchcmd{\doauthor}
  {\frontmatter@footnote{#2}}
  {#2}
  {}{\PackageError{authornotes}{Author-display patch failed}{}}
\makeatother

\begin{document}

\title{Field-Driven Active Matter Physics and Material Applications}

\author{Jing Wang}
\thanks{These authors contributed equally.}
\email{wangjing@ucas.ac.cn}
\affiliation{Wenzhou Institute, University of Chinese Academy of Sciences, Wenzhou, Zhejiang 325001, China}

\author{Shengkai Li}
\thanks{These authors contributed equally.}
\affiliation{School of Physics and Astronomy, Shanghai Jiao Tong University, Shanghai 200240, China}

\author{Trung V. Phan}
\thanks{These authors contributed equally.}
\affiliation{Department of Natural Sciences, Scripps and Pitzer Colleges, Claremont Colleges Consortium, Claremont, California 92110, USA}

\author{Gao Wang}
\affiliation{Wenzhou Institute, University of Chinese Academy of Sciences, Wenzhou, Zhejiang 325001, China}

\author{Huaicheng Chen}
\affiliation{Wenzhou Institute, University of Chinese Academy of Sciences, Wenzhou, Zhejiang 325001, China}

\author{Xiaorui Chen}
\affiliation{Department of Physics, School of Physical Science and Technology, Ningbo University, Ningbo, Zhejiang 315211, China}

\author{Robert H. Austin}
\email{austin@princeton.edu}
\affiliation{Department of Physics, Princeton University, Princeton, New Jersey 08544, USA}

\author{Liyu Liu}
\email{liuliyu@fudan.edu.cn}
\affiliation{Human Phenome Institute, Fudan University, Shanghai 201203, China}

\date{\today}

\begin{abstract}
Active matter consists of energy-consuming units that generate motion, making it a powerful platform for exploring nonequilibrium physics and engineering adaptive materials. Robotic active matter brings this physics into programmable, macroscopic agents with controllable actuation, sensing, and response, but its scalability is limited by onboard power, communication, and control complexity. Field-driven active matter addresses these limitations by shifting part of the control burden from individual agents to external or environmental fields, such as magnetic, electric, acoustic, optical, and fluidic fields, which can power, guide, and couple many agents simultaneously. Beyond applications in targeted navigation, adaptive collaboration, and flexible transport, field-driven robotic active matter offers a controllable testbed for nonequilibrium phenomena that are difficult to isolate in molecular or colloidal active matter, including tunable interactions, programmable phase transitions, solid–fluid switching, hyperuniform order, and reconfigurable collective motion. Here, we present field-driven active matter as a bridge between active matter physics and functional material engineering, organized through a unified \textit{field-to-information-to-physics-to-function} framework that links field-mediated interactions, information processing, and memory to collective transport, reconfiguration, and mechanical performance. By distributing control between agents and fields, field-driven active matter offers a pathway toward intelligent materials capable of sensing, adaptation, computation, reconfiguration, and performing complex tasks, while expanding both the scientific scope of nonequilibrium physics and the practical capabilities of engineered matter.
\end{abstract}

\maketitle
\thispagestyle{plain}

\tableofcontents

\vspace*{0.15in}

\section*{Overview}

\begin{figure}[t]
    \centering
\includegraphics[width=0.9\linewidth]{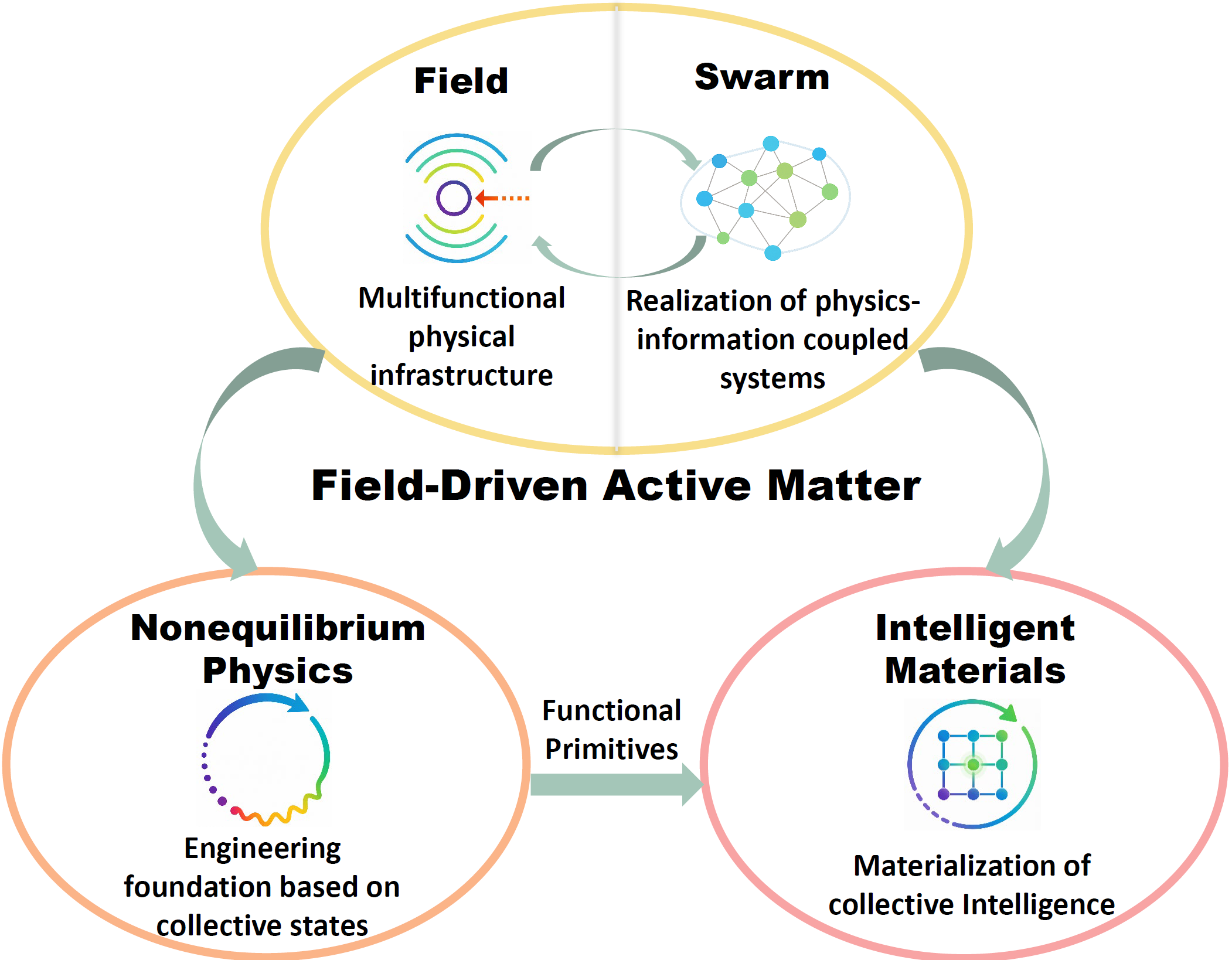}
\caption{\textbf{Schematic overview of this review.} }
\label{overview}
\end{figure}

Field-driven active matter describes active systems in which magnetic, electric, acoustic, optical, fluidic, or resource fields power, coordinate, and couple active units to exhibit nonequilibrium physical phenomena and form dynamic functional materials \cite{needleman2017active,han2025field,chen2022magnetic,wang2021emergent,wang2024robo}, as shown in Fig.~\ref{overview}. In these systems, fields do more than actuate agents. They can inject energy, mediate interactions, encode control signals, and, in adaptive environments, store information about past motion. A field can therefore become an \textit{active memory medium}: agents reshape the field, and the reshaped field feeds back on their future dynamics.

This review is organized as follows. Section \ref{sec:need} introduces field-driven active matter by classifying the interaction transmission range of active-matter systems\cite{han2025field,chen2022magnetic,wang2021emergent}. Section \ref{sec:fields} explains how different fields inject energy, mediate interactions, encode information, and act as memory media \cite{han2025field,chen2022magnetic,wang2021emergent,lavergne2019group}. Section \ref{sec:scales} surveys robotic platforms across scales \cite{yang2025magnetic,kemsaram2025acoustobots,xie2019reconfigurable,li2019particle}. Section \ref{sec:nonequi} discusses key nonequilibrium phenomena such as programmable phase transitions, hyperuniform order, memory-induced dynamics, and collective transport \cite{wang2025hyperuniform,wang2021emergent,lavergne2019group}. Section \ref{sec:Coupling} develops a unified physics-information coupling framework. Section \ref{sec:Material} maps these physics principles to material functions including adaptive mechanics, reconfiguration, transport, and self-healing \cite{wang2024robo,yang2025magnetic,xie2019reconfigurable}. Section \ref{sec:benchmarking} discusses benchmarking and quantification. Section \ref{sec:challenges} identifies deployment challenges and future directions. Section \ref{sec:conclusion} concludes with an outlook toward information-driven intelligent active matter.

Here, rather than focusing only on modular robotics or active matter separately, we establish a unified \textit{field-to-information-to-physics-to-function} framework that connects field-mediated nonequilibrium dynamics with information processing, memory, material performance, and application pathways. This perspective treats fields not only as actuation mechanisms, but also as media that encode information, mediate interactions, and shape functional collective states.

\section{\label{sec:need}From Active Matter to Field-Driven Active Matter}

\begin{figure*}[t]
    \centering
\includegraphics[width=1.0\linewidth]{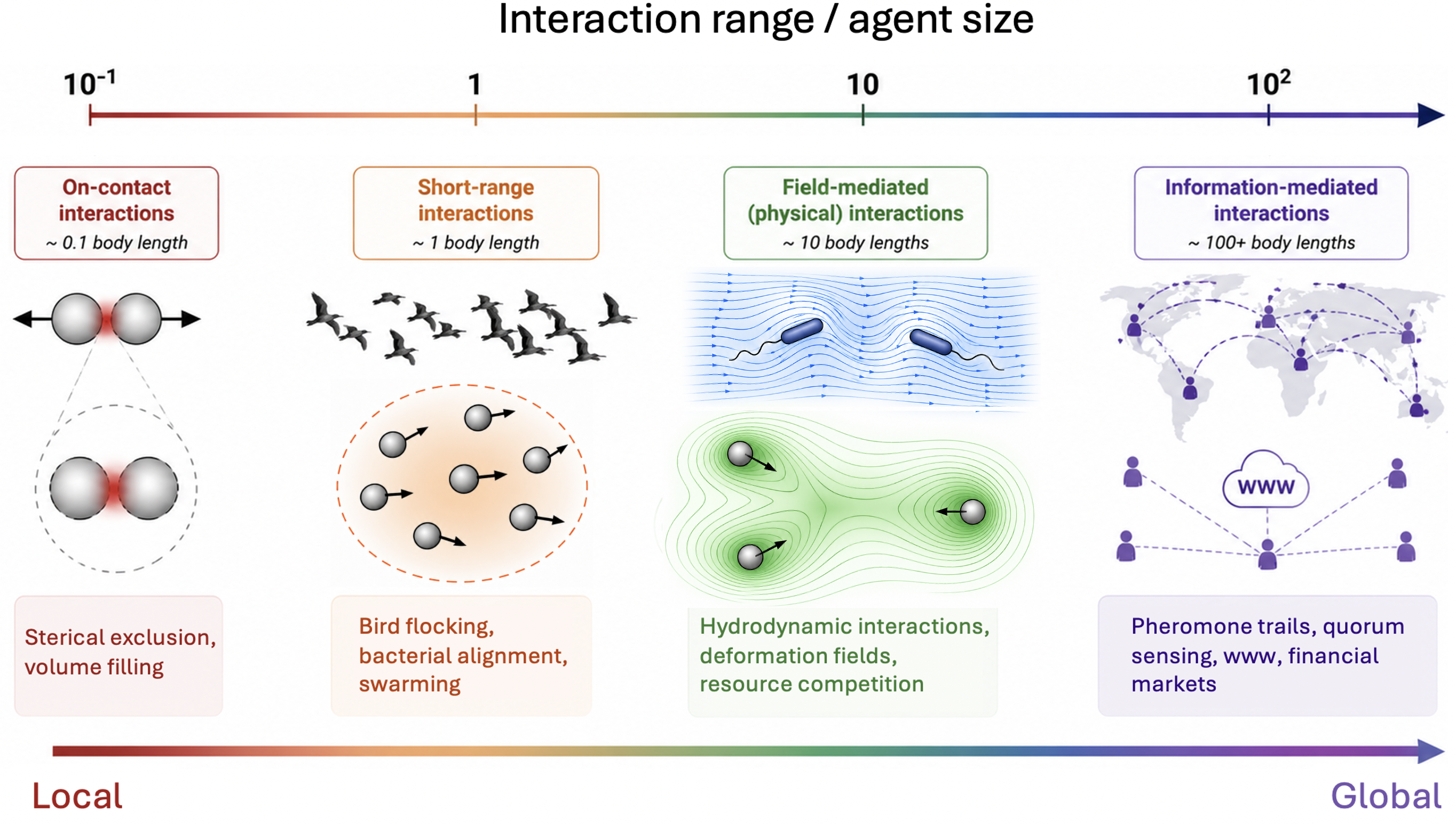}
\caption{\textbf{Interaction transmission range across active-matter systems.} Schematic comparison of active-matter systems according to the characteristic spatial range over which interactions are transmitted, normalized by the agent body length. From left to right, the representative transmission scale increases from local to global: on-contact interactions, where direct steric contact and excluded-volume interactions act within the immediate contact region; short-range interactions, where neighboring agents influence one another through local alignment and motion, as exemplified by bird flocking, bacterial alignment, and swarming; field-mediated physical interactions, where spatially extended physical fields transmit interactions between separated agents, illustrated by hydrodynamic flow lines, deformation or resource fields, and the associated agent responses; and information-mediated interactions, where interactions are transmitted through communication or information networks over distances much larger than the agent size, as illustrated by pheromone trails, quorum sensing, the World Wide Web, and financial networks. Arrows indicate agent motion or orientation where shown, while colored field lines and contours represent spatially extended physical fields. The numerical positions indicate representative order-of-magnitude transmission scales rather than universal boundaries between interaction regimes. The regimes are not mutually exclusive, and a given system may involve multiple interaction mechanisms operating at different spatial scales.}
\label{IntRange}
\end{figure*} 

Active matter is broadly composed of energy-consuming units that continuously convert energy into directed motion or mechanical forces, thereby sustaining nonequilibrium dynamics and enabling collective behavior\cite{needleman2017active}. In this review, we adopt a functional criterion for defining the scope of field-driven active matter rather than restricting it to a particular material, length scale, or actuation mechanism. We include systems when (i) their constituent agents are active and sustain nonequilibrium motion or force generation through energy consumption, (ii) an external or environmental field plays a causal role in governing agent motion, interactions, coordination, or information exchange, and (iii) field-mediated collective dynamics are central to the physical behavior or function of the system. This criterion encompasses field-driven active colloids, microswarms, macroscale robotic collectives, and active material systems, including self-healing systems when their recovery arises from field-driven collective reorganization rather than passive material response\cite{massana2019tunable,han2020reconfigurable}. Conversely, passive field-responsive materials without active agents or field-mediated collective dynamics are outside the primary scope of this review.

Interactions are the central mechanism through which active matter converts individual activity into collective behavior. Classical studies of active matter focused primarily on self-propulsion, energy injection, and dissipation, demonstrating that even simple active units can exhibit flocking, clustering, phase separation, and collective transport \cite{vicsek1995novel,ramaswamy2010mechanics,marchetti2013hydrodynamics,bechinger2016active}. As the field matured, increasing attention was devoted not only to how active agents move, but also to how they influence one another. From this perspective, active-matter systems can be viewed as occupying a spectrum defined by the spatial range over which interactions are transmitted, as shown in Fig.\ref{IntRange}.

At one extreme are contact-dominated systems, where interactions occur only when agents physically collide. Examples include steric exclusion in colloidal suspensions, granular active matter, and densely packed robotic collectives. The effective interaction range is typically confined to the immediate contact region and is therefore represented in Fig.\ref{IntRange} by a characteristic scale of approximately $10^{-1}$ agent body lengths. In such systems, information about neighboring agents is transmitted almost exclusively through direct mechanical contact. Despite this highly local nature, contact interactions alone can generate collective phenomena including jamming, clustering, and mechanically induced phase transitions \cite{li2021programming,saintyves2024self}.

Increasing the interaction range to the scale of the agent body length allows agents to respond to nearby neighbors without direct contact, dramatically enriching collective behavior. Alignment-based systems such as bird flocks, fish schools, bacterial collectives, and the Vicsek model therefore occupy the $\sim$ 1 body-length regime\cite{vicsek1995novel,deseigne2010collective,bricard2015emergent}. Here these interaction transmission length values should be understood as representative order-of-magnitude scales rather than universal boundaries; the effective range depends on the specific interaction law, agent geometry, and physical environment. Although still local, these interactions allow orientational information to propagate through the population, producing large-scale flocking, swarming, and coherent collective motion. The emergence of order in such systems illustrates how modest increases in interaction range can profoundly alter the macroscopic phase behavior of active matter.

A further extension occurs when interactions are mediated by physical fields. Hydrodynamic flows generated by swimming microorganisms, deformation fields created in elastic substrates, capillary interactions at interfaces, and chemical concentration fields surrounding active particles all allow agents to influence one another over distances substantially larger than their own size \cite{drescher2009dancing,saintillan2013active,theurkauff2012dynamic,palacci2013living,nishiguchi2018engineering}. For the representative systems considered here, the characteristic transmission distance is commonly on the order of several to tens of agent body lengths, and Fig.\ref{IntRange} uses $\sim 10$ body lengths as a representative scale. This value is not intended as a universal threshold; rather, it provides an order-of-magnitude reference for distinguishing spatially extended field-mediated coupling from predominantly local interactions. In this regime, the environment becomes an active participant in the dynamics. Agents modify a field; the field propagates through space, and other agents respond to the resulting environmental changes. Such field-mediated coupling introduces effective long-range interactions that can generate vortices, collective transport, synchronization, and emergent material properties \cite{wang2021emergent,phan2021bootstrapped}.

The largest interaction ranges arise when fields function not only as physical mediators but also as carriers of information, such as when interactions are mediated by communication networks, persistent environmental signals, or information-processing architectures. Human societies connected through the World Wide Web, communication-enabled robotic swarms, and biological systems employing persistent chemical or environmental memory can transmit information over distances many orders of magnitude larger than the physical size of individual agents\cite{jackson2004trail,papenfort2016quorum}. Accordingly, Fig.\ref{IntRange} places information-mediated interactions in a representative $100+$ body-length regime. Here again, the numerical value is intended only as an order-of-magnitude reference, rather than a threshold separating information-mediated from physical-field-mediated interactions. In these systems, fields no longer merely transmit forces; they encode, store, and distribute information. Interaction becomes increasingly detached from physical proximity and increasingly governed by communication networks, memory traces, and information-processing architectures. Information fields therefore represent a natural extension of field-mediated active matter toward intelligent collective systems.

\begin{figure*}[!ht]
    \centering
\includegraphics[width=1.0\linewidth]{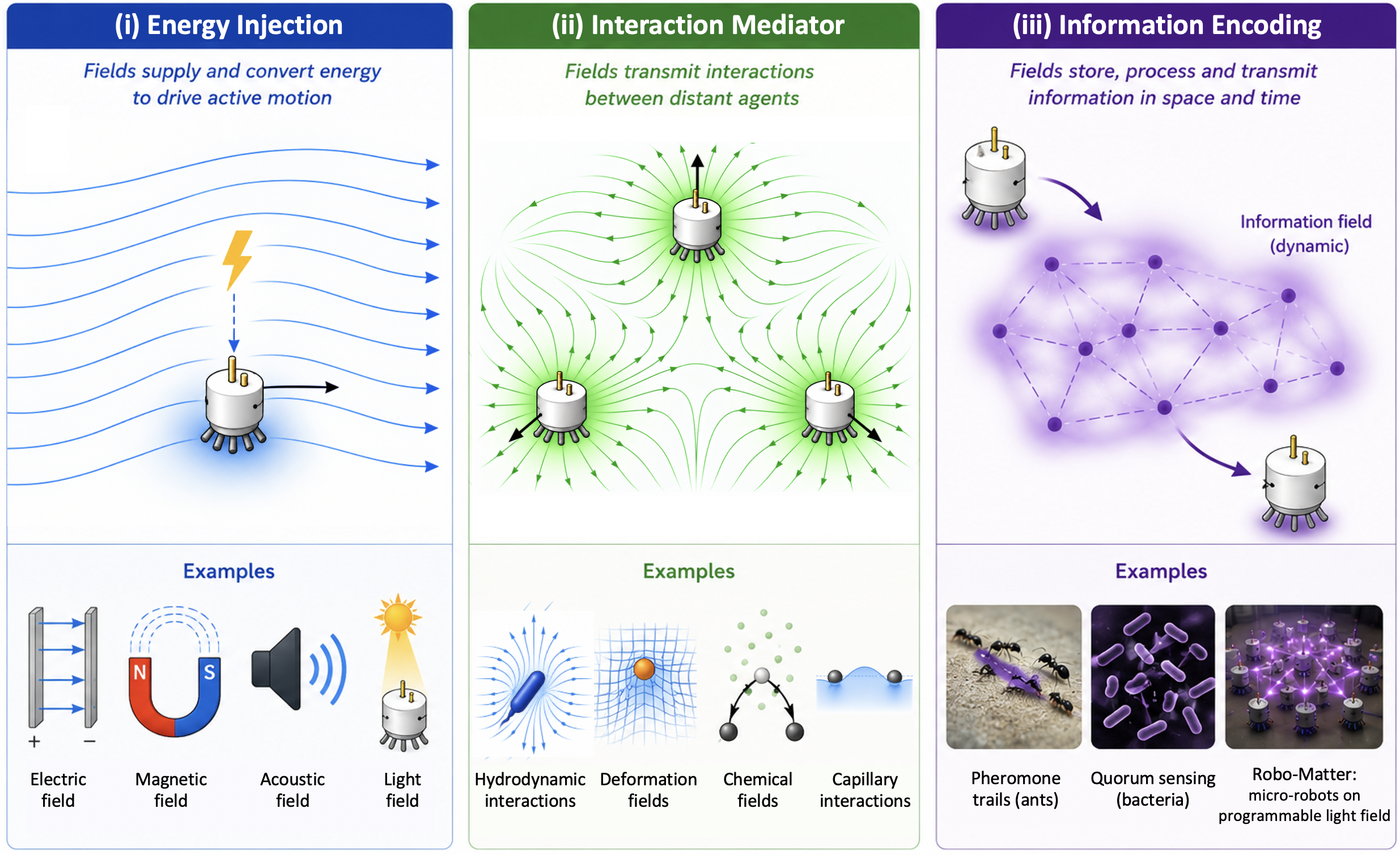}
\caption{\textbf{Roles of fields in active matter.} Schematic illustration of three distinct but potentially overlapping roles of physical fields in active-matter systems. (i) Energy injection: external fields supply or convert energy to drive active motion. The blue field lines represent the applied field, the lightning symbol indicates energy transfer from the field to the active agent, and the black arrow indicates the resulting direction of agent motion. Representative examples include electric, magnetic, acoustic, and light fields. (ii) Interaction mediator: fields transmit physical interactions between spatially separated agents. The green field lines and arrows represent the spatial distribution and direction of the mediating field, while the black arrows indicate the corresponding agent responses. Examples include hydrodynamic interactions, deformation fields, chemical fields, and capillary interactions. (iii) Information encoding: fields can spatially and temporally store, process, and transmit information that regulates agent behavior. The purple shaded region represents a dynamic information field, the nodes and dashed links represent information-mediated connectivity, and the purple arrows indicate information exchange between agents and the field. Representative examples include pheromone trails, quorum sensing, and programmable light fields in Robo-Matter\cite{jackson2004trail,papenfort2016quorum,wang2024robo}. These roles are not mutually exclusive, and a field need not serve as the primary energy source of the active agents.}
\label{roles}
\end{figure*} 

Viewed through this lens, the four regimes in Fig.\ref{IntRange} provide a spatial-transmission perspective for comparing how interactions propagate through active-matter systems, rather than a developmental hierarchy. Contact forces, neighbor-based alignment, field-mediated coupling, and information-mediated communication represent distinct mechanisms that can operate at different characteristic ranges. Importantly, these mechanisms are not mutually exclusive: a single active-matter system may simultaneously involve contact interactions, short-range alignment, field-mediated coupling, and information exchange, with different mechanisms serving different physical or functional purposes. The numerical positions in Fig.\ref{IntRange} therefore indicate representative transmission scales rather than universal classification boundaries or levels of sophistication. This perspective unifies diverse active-matter systems within a common framework and highlights the importance of fields as mechanisms that amplify interaction range, enable environmental feedback, and ultimately allow collective systems to process information. Field-driven robotic active matter occupies a particularly important position in this framework because it provides a programmable platform in which the range, strength, and informational content of interactions can be engineered directly. As a result, it serves as a bridge between conventional active matter and future information-driven active matter, where collective behavior emerges not only from the flow of energy and matter, but also from the flow of information itself.

It is important to emphasize that the classifications adopted in this review describe complementary, rather than mutually exclusive, dimensions of field-driven active matter. The interaction-range framework in Section \ref{sec:need} characterizes how interactions are transmitted across a system, from direct contact to field- and information-mediated coupling. Section \ref{sec:fields} instead asks what functional role the field plays, distinguishing energy injection, interaction mediation, and information encoding. Section \ref{sec:scales} then considers how these principles are embodied in robotic collectives across length scales and levels of distributed computation. Sections \ref{sec:nonequi}–\ref{sec:Material} progressively address what collective physics emerges, how physics and information co-evolve, and how these dynamics are translated into functions and material properties. Thus, these categories should not be interpreted as separate classes of systems, but as different projections of the same underlying field–agent–environment system, linking interaction mechanism, physical realization, emergent dynamics, and material function.

This cross-scale scope distinguishes the present review from existing reviews that have primarily focused on specific classes of field-driven systems, such as field-driven colloids, magnetic active matter, microswarms, or robotic matter. Rather than treating these systems as separate research domains, we use a common framework to connect them through the transmission mechanism and functional role of fields. In particular, we identify a progression from field-mediated interactions to information processing, collective physics, functional primitives, and emergent material properties, summarized by the pathway Field to Information to Physics to Function to Material Property. This perspective highlights that fields can serve not only as sources of energy, but also as physical media for interaction mediation, information encoding, and memory, while robotic collectives provide an experimentally accessible bridge between microscopic active matter and functional intelligent materials. Building on this framework, this review further proposes quantitative benchmarks for comparing collective physical performance, information-enabled functions, and emergent material properties across otherwise disparate platforms.

{\section{\label{sec:fields}Fields as Multiple Roles: From Energy Injection to Information Encoding}}

In the conventional systems of active matter, external driving fields, such as magnetic, electric, acoustic, optical, and hydrodynamic fields, are primarily regarded as external actuation mechanisms that inject energy into individual agents and thereby sustain nonequilibrium dynamics\cite{marchetti2013hydrodynamics}. Within this traditional perspective, fields function mainly as sources of forces and torques, while collective behaviors emerge from the interactions among active agents themselves\cite{anderson1972more}. The emergence of field-driven active matter has substantially expanded this view. As active matter increasingly intersects with robotics, programmable materials, and embodied intelligence, external fields are evolving from simple energy-delivery mechanisms into multifunctional physical infrastructures\cite{bechinger2016active,ju2025technology}. Beyond powering individual agents, fields can dynamically regulate interactions, coordinate collective behaviors, and encode task-relevant information directly into the physical environment, as shown in Fig.\ref{roles}. Consequently, collective dynamics become shaped not only by energetic constraints but also by the way information is embedded, transmitted, and processed through field-mediated interactions. This evolution reflects a broader conceptual transition: from controlling individual particles to programming collective behavior, and ultimately toward integrating physical dynamics and information processing within a unified framework.

In this review, field-driven refers not exclusively to the energetic origin of activity, but more broadly to the dynamical role of external or environmental fields in governing active systems. Specifically, field-driven systems here include field-powered systems, in which the field directly supplies energy for active motion, and field-mediated or field-controlled systems, in which the active agents may possess internal energy sources while the field regulates their interactions, motion, coordination, or information exchange. Thus, a macroscopic robot equipped with an onboard battery can still be considered field-driven when an external field plays a central role in determining its collective dynamics. This broader definition is motivated by the central premise of this review: fields provide a physical infrastructure through which control can be distributed from individual agents to the collective. Accordingly, the three roles considered here—energy injection, interaction mediation, and information encoding—represent distinct but potentially overlapping ways in which fields govern nonequilibrium dynamics. The term field-driven active matter therefore emphasizes field-governed dynamics rather than field-supplied energy alone. This distinction also clarifies the scope of the review and allows field-powered microswimmers and internally powered robotic collectives to be discussed within a common physics–information framework.\\

\subsection{\label{subsec:hist}Historical Evolution of Field-Driven Systems}

The role of external fields in active matter has undergone a progressive transformation over the past several decades. This evolution can be broadly understood through three conceptual stages. The first stage viewed fields primarily as sources of energy and actuation. Early studies of ferrofluids, colloidal suspensions, and electrokinetic systems employed static magnetic or electric fields to induce particle alignment, aggregation, and transport\cite{scherer2005ferrofluids,winslow1949induced,trau1996field,grzybowski2002dynamic}. A major technological milestone was the invention of optical tweezers by Ashkin in the 1980s\cite{ashkin1987optical,ashkin1987opticalcells}, which enabled precise manipulation of microscopic objects and established external fields as powerful tools for single-particle control. The second stage emerged with the rise of active matter physics, where fields became regulators of collective interactions rather than merely providers of energy. A landmark example was the demonstration of large-scale collective motion in Quincke rollers by Bricard \textit{et al.} \cite{bricard2013emergence}, where a uniform electric field generated self-propulsion and collective flocking-like dynamics. Similar developments in magnetic colloids, active droplets, and self-propelled particles revealed that external fields could reshape effective interaction networks, stabilize nonequilibrium states, and induce collective phase transitions\cite{palacci2013living,snezhko2011magnetic,takatori2014swim,MattSullivan2003,sullivan2006electric}. More recently, a third stage has begun to emerge in field-driven active matter, where fields function not only as sources of energy or interaction mediators but also as carriers of information. Through programmable spatial and temporal modulation, control policies can be encoded directly into field landscapes and translated into collective responses without requiring complex onboard computation\cite{sitti2021physical,marzo2015holographic,yu2019active,melde2016holograms}. As a result, collective behavior becomes shaped by both physical interactions and information embedded within the surrounding environment. These stages reflect an increasing level of functional abstraction rather than purely chronological development. This progression—from energy injection, to interaction regulation, to information encoding—represents a fundamental conceptual transformation in the design of active systems. It establishes the conceptual foundation for understanding robotic active matter as a field-mediated physics–information system.

\subsection{\label{subsec:energy}Fields as Energy Injection Mechanisms}

\begin{figure}[ht]
    \centering
\includegraphics[width=1.0\linewidth]{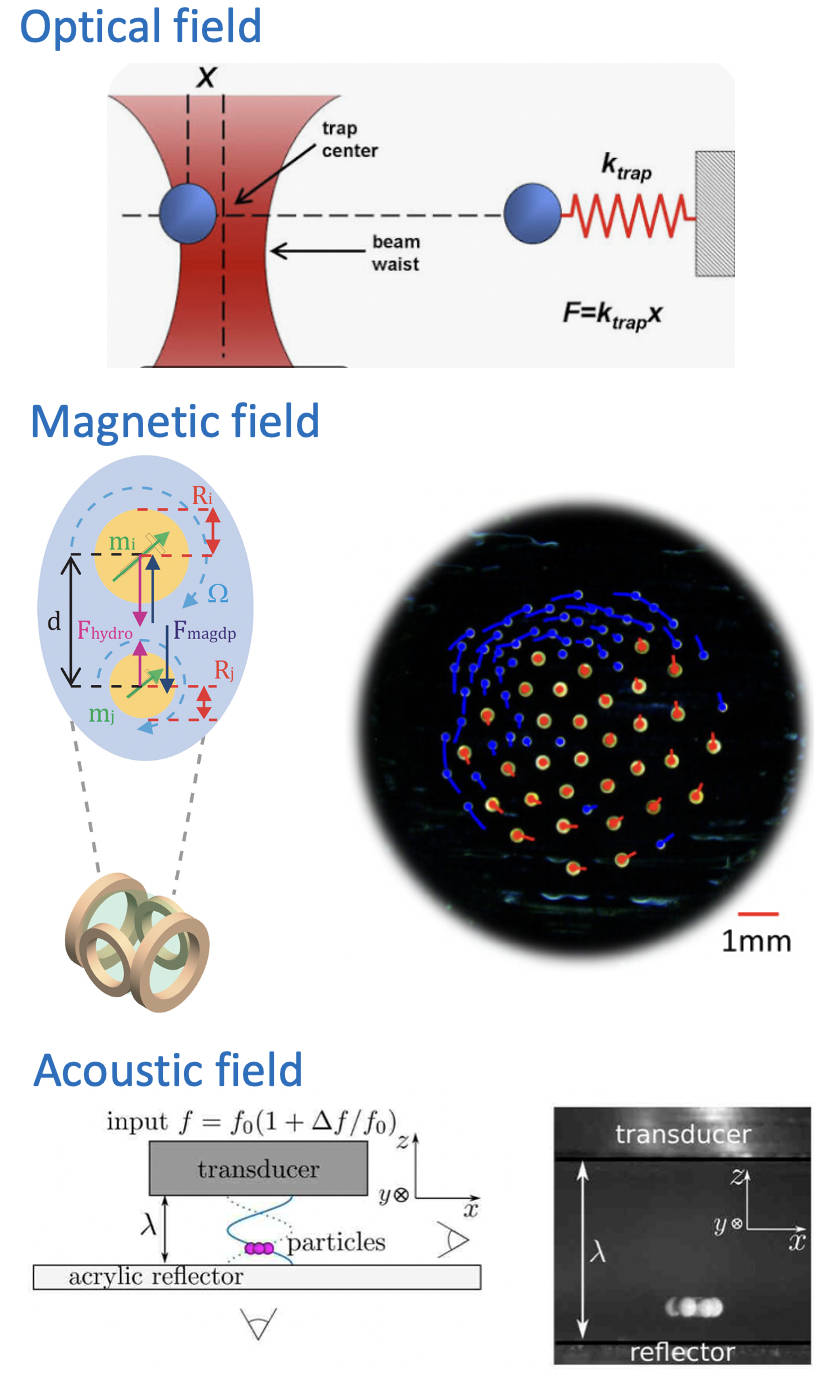}
\caption{\textbf{Field as energy injection.} External fields provide energy to drive active motion. Representative examples include optical trapping, magnetic actuation, and acoustic manipulation, illustrating how different energy fields can power and control active particles and microrobots\cite{moffitt2006differential,ceron2023programmable,lim2019cluster}. Top panel originally created by RockyRaccoon, public domain, via Wikimedia Commons \cite{WikimediaOpticalTrapSpring}. Middle panel reproduced from Ceron \textit{et al.}, Proc. Natl. Acad. Sci. U.S.A. \textbf{120}(24), e2221913120 (2023); licensed under a Creative Commons Attribution license 4.0 (CC BY) . Bottom panel reproduced with permission from Lim \textit{et al.}, Nat. Phys. \textbf{15}(5), 460-464 (2019). Copyright 2019 Springer Nature.}
\label{energy}
\end{figure}

A defining feature of active matter is the continuous conversion of energy into directed motion or forces at the level of individual agents. Sustained energy injection is therefore a prerequisite for maintaining nonequilibrium dynamics\cite{bechinger2016active,bowick2022symmetry}. Depending on system scale and physical environment, ranging from low-Reynolds-number microscale systems to inertia-dominated macroscopic robots, the mechanisms of field-induced energy transduction differ significantly\cite{ning2024macroscopic}. External fields convert macroscopic energy inputs (optical, magnetic, electric, or acoustic) into localized kinetic or thermal energy, thereby powering active motion\cite{essafri2022designing}. 

Optical energy injection in active colloids is typically achieved through two distinct mechanisms: photocatalytic and photothermal propulsion. In photocatalytic systems, Janus particles (e.g., Ag-based) use light to trigger asymmetric chemical reactions (such as the decomposition of H$_2$O$_2$), generating concentration gradients that drive motion via self-phoretic effects\cite{hong2010light,dong2018photocatalytic,howse2007self}. In contrast, photothermal mechanisms rely on light absorption to create local temperature gradients, inducing thermophoretic flows or convection that propel particles\cite{jiang2010active,volpe2011microswimmers}. These represent two distinct pathways for converting optical energy into chemical potential gradients or thermal fields. In addition, as shown in the top panel of Fig.\ref{energy}, optical trapping forces and radiation pressure can directly manipulate particles\cite{ashkin1986observation,moffitt2006differential,WikimediaOpticalTrapSpring}. A key advantage of optical fields is their high spatial resolution, enabling precise control over local activity via programmable light landscapes\cite{palacci2013living,lozano2016phototaxis}.

Magnetic fields are widely used due to their deep penetration, non-invasive nature, and precise torque control. Compared to optical fields, which are strongly scattered in biological tissues, and acoustic fields, which are limited by impedance matching, magnetic fields offer clear advantages for \textit{in vivo} and deep-tissue applications\cite{nelson2010microrobots,sitti2015biomedical}. The relative importance of torque- versus force-dominated regimes depends on system size and frictional environment. At low Reynolds numbers, energy is injected via magnetic torque, $\tau = m \times B$ which drives rotational or undulatory motion against viscous resistance\cite{zhang2009artificial,ghosh2009controlled}. At larger scales, magnetic gradient forces, $F = \nabla(m \cdot B)$ enable direct translational actuation, with force outputs spanning from piconewtons (pN) to newtons (N)\cite{kummer2010octomag}. For instance, as shown in the middle panel of Fig.\ref{energy}, rotating magnetic fields can induce particle rotation and generate viscous flow\cite{xie2019reconfigurable,ceron2023programmable}; symmetry breaking near boundaries can lead to net fluid transport\cite{tierno2008controlled}; and time-modulated magnetization enables artificial cilia for fluid pumping\cite{ul2022microscopic,den2008artificial}. A key advantage of magnetic actuation lies in its global and scalable energy delivery, enabling synchronized driving of large populations without increasing onboard complexity.

Electric-field-driven systems rely on interfacial electrohydrodynamic effects such as Quincke rotation and induced-charge electro-osmosis (ICEO)\cite{melcher1969electrohydrodynamics,bazant2004induced}. Dielectric particles subjected to a DC electric field can undergo spontaneous Quincke rotation, leading to rolling motion and sustained propulsion\cite{bricard2013emergence,bricard2015emergent}. When the field exceeds a threshold, rotation frequencies can reach hundreds of Hz, generating electrohydrodynamic flows that inject energy into the system\cite{pannacci2007rheology,lemaire2008viscosity}. Temporal modulation of the field (e.g., pulsed driving) enables control over particle velocity and collective coherence time\cite{karani2019tuning,pradillo2019quincke}. ICEO-driven Janus particles exploit asymmetric polarization under AC fields, generating tangential slip flows that propel particles with a nonlinear scaling $v \propto E^2$, offering rich control possibilities\cite{gangwal2008induced,squires2004induced}.

Acoustic fields drive particles through acoustic radiation forces and acoustic streaming\cite{bruus2012acoustofluidics,bruus2011forthcoming,mcneill2023acoustically,lim2019cluster}, as shown in the bottom panel of Fig.\ref{energy}. Structured acoustic cavities can generate acoustic potential wells that trap particles\cite{huang2026sound}, while asymmetric geometries produce streaming flows that propel micromotors\cite{nadal2014asymmetric}, and phased transducer arrays enable three-dimensional levitation and transport\cite{marzo2015holographic}. At larger scales, acoustic robots (e.g., AcoustoBots\cite{kemsaram2025acoustobots}) operate under the combined influence of streaming and inertial effects. In addition, hydrodynamic fields, such as externally imposed shear flows or microfluidic forcing, can also drive particle alignment, transport, and pattern formation\cite{rusconi2014bacterial,saintillan2008instabilities,haghighat2016characterization}.

Unlike internally powered active systems, field-driven systems externalize the energy source. This separation enables precise global control, scalability, and programmability. Importantly, energy injection is often spatially and temporally heterogeneous: gradients, modulation, and field topology introduce anisotropy, which plays a crucial role in generating complex collective behaviors and nonequilibrium phase transitions\cite{khadem2021transport}.

\subsection{\label{subsec:InterMed}Fields as Interaction Mediators}

\begin{figure}[t]
    \centering
\includegraphics[width=0.9\linewidth]{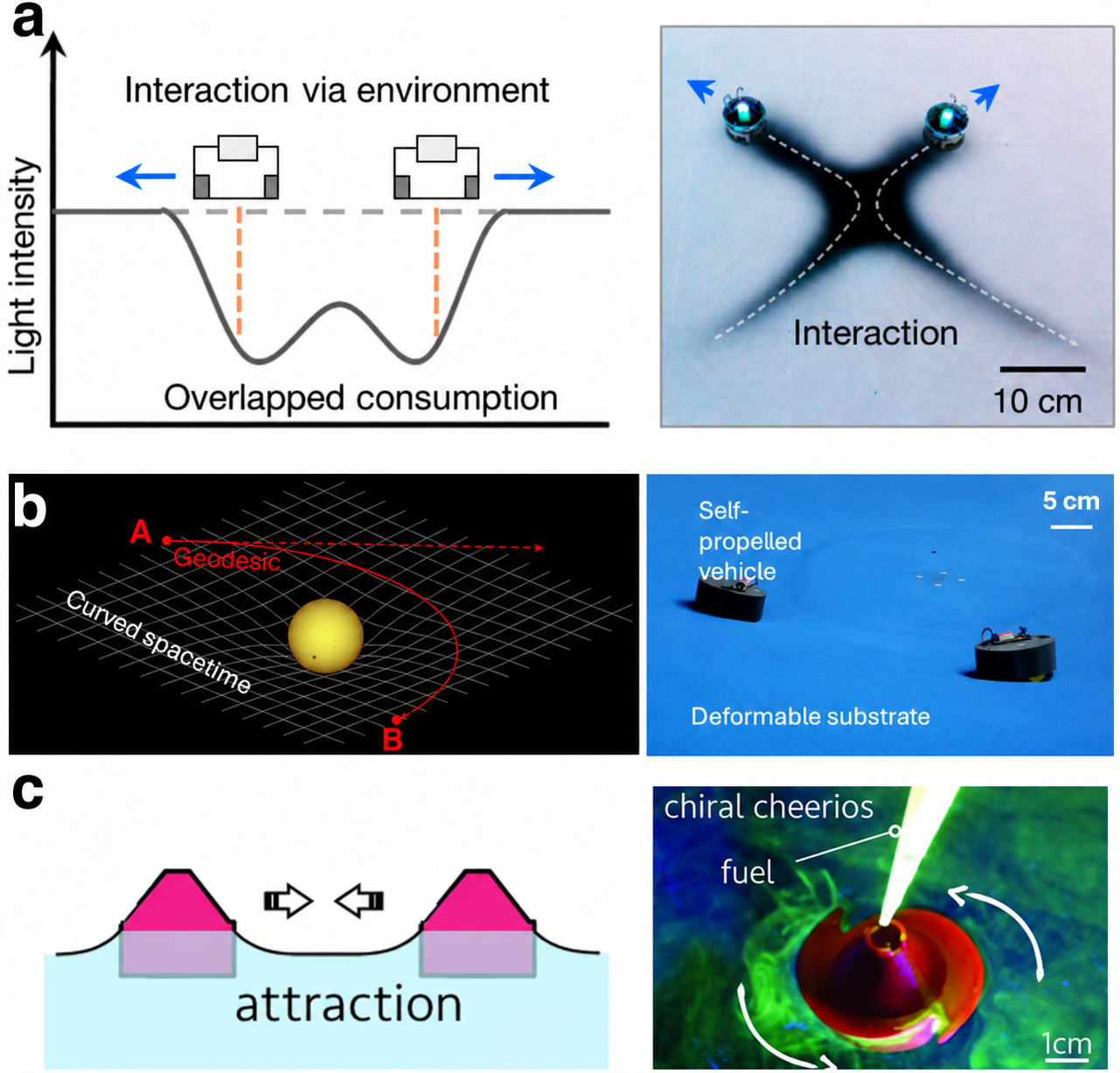}
\caption{\textbf{Field as interaction mediator.} Agent interactions mediated by shared environmental fields: (a) Robots repel via overlapping depletion of a common light field\cite{wang2022robots}. Reproduced from Wang \textit{et al.}, Proc. Natl. Acad. Sci. U.S.A. \textbf{119}(12), e2120019119 (2022); licensed under a Creative Commons Attribution-NonCommercial-NoDerivatives License 4.0 (CC BY-NC-ND). (b) Analogous to geodesics in curved spacetime, self-propelled vehicles couple mechanically through deformable substrate curvature\cite{li2022field}. Reproduced from Li \textit{et al.}, Proc. Natl. Acad. Sci. U.S.A. \textbf{119}(30), e2113912119 (2022); licensed under a Creative Commons
Attribution-NonCommercial-NoDerivatives License 4.0 (CC BY-NC-ND). (c) Interfacial capillary fields attract floating chiral Cheerio swimmers. The graph on the right shows fuel-driven chiral particle rotation at the fluid surface\cite{wilt2024activecheerios}. Reproduced from Wilt \textit{et al.}, arXiv:2411.16011 (2024); licensed under a Creative Commons Attribution License 4.0 (CC BY).}
\label{mediator}
\end{figure}

While energy injection drives individual particles away from equilibrium, interactions determine how this activity propagates throughout the system. Collective phenomena in active matter do not arise solely from the self-propulsion of individual agents, but from the coupling mechanisms that transmit mechanical influences across many-body assemblies. In field-driven active matter, external fields can play a dual role: they not only provide energy input but also mediate interactions between particles, thereby defining the effective communication channels through which collective dynamics emerge\cite{snezhko2011magnetic,bricard2013emergence}. These interaction mechanisms can be classified according to their spatial range, physical origin, and symmetry properties.

At the most fundamental level, field-mediated interactions originate from pairwise couplings between neighboring particles. Depending on the driving modality, these interactions may arise from magnetic dipolar forces, electrostatic interactions, acoustic radiation forces, optical forces, or other externally imposed potentials\cite{palagi2018bioinspired,bose2022collective}. Such pairwise interactions are often anisotropic and long-ranged, allowing particles to attract, repel, align, or assemble into ordered structures\cite{liljestrom2019active,bechinger2016active}. For example, rotating magnetic fields can induce dipolar attractions that promote chain formation, clustering, or lattice-like organization\cite{xie2019reconfigurable,jin2021collective,yu2019active}. Beyond direct force-mediated interactions, active agents can also modify their surrounding physical fields and thereby influence neighboring agents, as in active matter systems, where interactions are mediated by shared environmental fields—namely light fields\cite{wang2022robots}, elastic fields\cite{li2022field}, or hydrodynamic fields\cite{wilt2024activecheerios}, as shown in Fig. \ref{mediator}. It is worth noting that field-mediated coupling may operate through two distinct mechanisms: mechanical interactions, where fields exert physical forces, and informational interactions, where fields encode environmental cues that guide agent-level decisions. In the latter case, collective dynamics emerge from distributed information processing rather than direct force transmission. A representative example of field-mediated interactions is the interaction between two walking droplets, where self-generated surface pilot waves mediate effective forces and energy exchange between walkers, leading to nontrivial pairwise attraction or repulsion behaviors\cite{borghesi2014interaction,damiani2026energetics,8z1kc144}. These elementary interaction rules establish the basic physical framework from which larger-scale collective behavior emerges. 

However, many active systems cannot be understood solely in terms of pairwise interactions. Particle motion frequently generates secondary physical fields that propagate through the surrounding medium, giving rise to long-range collective coupling\cite{marchetti2013hydrodynamics}. Elastic deformations transmitted through compliant substrates and boundaries\cite{li2022field,ghosh2016interactions,deblais2018boundaries,giomi2013swarming} (see Fig. \ref{mediator}(b)), hydrodynamic flows generated by rotating particles\cite{zhang2020reconfigurable,wilt2024activecheerios} (see Fig. \ref{mediator}(c)), and concentration gradients produced by catalytic reactions\cite{popescu2016self} can all act as interaction mediators. In such cases, the behavior of a given particle depends not only on its immediate neighbors but also on collective modifications of the surrounding environment. These field-mediated couplings effectively create nonlocal interaction networks, enabling synchronization, collective transport, and coherent swarm-level dynamics over distances far exceeding individual particle dimensions.

A distinctive feature of many field-driven active systems is the emergence of nonreciprocal interactions\cite{fruchart2021non}. In equilibrium systems, interactions generally obey action–reaction symmetry, such that the influence exerted by one particle on another is balanced by an equal and opposite response. Active systems, by contrast, often operate through delayed, field-dependent, or self-generated interaction pathways that break this symmetry. As a result, the effective influence of particle A on particle B may differ from that of particle B on particle A. Nonreciprocal interactions naturally arise in chemically mediated systems, hydrodynamically coupled rotors, and field-driven robotic collectives, where sensing, actuation, and environmental feedback are intrinsically intertwined\cite{soto2014self,wu2021ion,agudo2019active,soni2019odd,chen2024emergent}. Such asymmetries constitute a defining hallmark of nonequilibrium matter and underlie a broad range of active phenomena that have no equilibrium counterpart.

The most significant distinction between field-driven active matter and conventional active matter lies in the programmability of these interaction networks\cite{yu2019active,xie2019reconfigurable}. In classical active systems, interaction rules are largely fixed by particle geometry, material properties, and environmental conditions. By contrast, external fields allow interaction strengths, ranges, symmetries, and even network topologies to be dynamically reconfigured in real time\cite{lavergne2019group}. Adjusting magnetic frequencies, acoustic phases, optical intensities, or electric-field configurations can continuously reshape the effective coupling network without modifying the particles themselves\cite{han2024imaging,katuri2022motility,collins2018self,juarez2012feedback,fernandez2020feedback}. Consequently, a single experimental platform can access multiple collective states under different driving protocols, transforming interactions from passive physical constraints into actively controllable design parameters. This transition from fixed interactions to programmable interaction networks represents a crucial step toward adaptive active materials and naturally motivates the next question: how can information itself be encoded, stored, and processed within such dynamically reconfigurable systems?

\subsection{\label{subsec:InforCar}Fields as Information Carriers}

\begin{figure}[t]
    \centering
\includegraphics[width=1.0\linewidth]{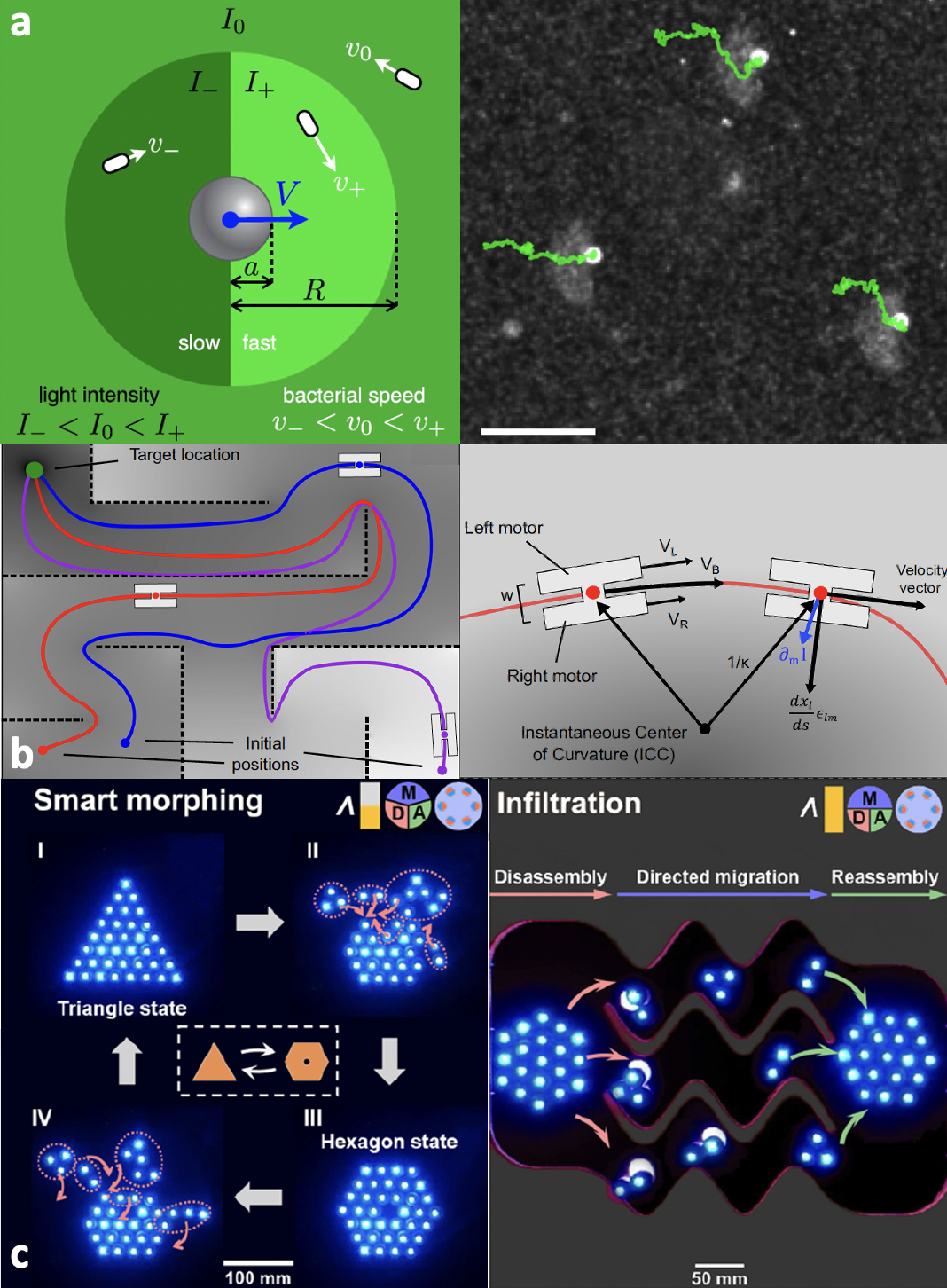}
\caption{\textbf{Field as information carrier.} Environmental fields can store, process, and transmit information to coordinate active agents. (a) Light-controlled bacterial activity gradients generate active-pressure fields that transport passive colloids\cite{pellicciotta2023colloidal}. Reproduced from Pellicciotta \textit{et al.}, Nat. Commun. \textbf{14}(1), 4191 (2023); licensed under a Creative Commons Attribution License 4.0 (CC BY). (b) Programmable sensory fields guide reactive robot navigation\cite{reinhardt2025artificial}. Reproduced from Reinhardt \textit{et al.}, npj. Robot. \textbf{3}(1), 39 (2025); licensed under a Creative Commons Attribution License 4.0 (CC BY). (c) Dynamic light fields coordinate Robo-Matter, enabling collective functions such as smart morphing and infiltration\cite{wang2024robo}. Reproduced from Wang \textit{et al.}, Nat. Commun. \textbf{15}(1), 8853 (2024); licensed under a Creative Commons Attribution License 4.0 (CC BY).}
\label{infocarrier}
\end{figure}

Beyond supplying energy and mediating interactions, external fields can also function as carriers of information. In this role, fields provide spatiotemporally structured signals that encode environmental constraints, task objectives, and collective coordination cues\cite{pellicciotta2023colloidal,reinhardt2025artificial,wang2021emergent,wang2022robots,wang2024robo}, as shown in Fig. \ref{infocarrier}. Rather than relying solely on onboard sensing, communication, and computation, field-driven active systems can exploit externally imposed information landscapes to guide individual and collective behavior\cite{ceylan2017mobile}. This capability fundamentally distinguishes field-driven active matter from conventional active systems, in which information processing is typically localized at the particle level.

At the most basic level, external fields encode information through their spatial and temporal structure. Gradients of light intensity, magnetic-field orientation, acoustic pressure, or electric potential define information landscapes, such as distributions of resources, pheromones, and obstacles, that particles can locally sample and respond to\cite{wang2024robo,heuthe2024counterfactual,salman2020phormica}. In this framework, navigation, targeting, and collective organization emerge from interactions with the field itself rather than from explicit communication among agents\cite{reinhardt2025artificial,yuan2024controlling,werfel2014designing}. For example, as shown in Fig. \ref{infocarrier}(a), the spatial intensity distribution of the light field can encode local bacterial swimming speed via proteorhodopsin-controlled motility, and collective density–polarization fields transduce this speed pattern into an active pressure gradient that drives passive colloids in controlled directions\cite{pellicciotta2023colloidal}. Information is therefore embedded directly into the physical environment, transforming space into a distributed computational substrate.

External fields also provide an efficient mechanism for large-scale information distribution. Unlike conventional robotic swarms, where information must be exchanged through pairwise communication links, field-driven systems can broadcast control signals simultaneously to large populations of agents\cite{brambilla2013swarm,wang2021emergent,kummer2010octomag}. A single global field can coordinate thousands of particles in parallel, enabling scalable collective behavior without requiring complex communication architectures\cite{ceron2023programmable,yu2019active}. For example, as shown in Fig. \ref{infocarrier}(b), through conformal transformation and virtual-space metric encoding, a static light field with spatially varying intensity becomes an artificial spacetime carrying both the structural information of the physical environment and the navigation control policy, thereby enabling light-driven differential-drive microrobots without onboard computation or communication to perform decentralized parallel navigation, obstacle avoidance, and turning\cite{reinhardt2025artificial}. This property is particularly advantageous at micro- and nanoscales, where onboard communication and computation remain severely constrained by size, power, and fabrication limitations.

More importantly, fields can externalize part of the computational burden traditionally assigned to individual agents. In many field-guided systems, the environment itself performs a form of physical computation by embedding task-relevant information into the field configuration\cite{wang2024robo,salman2020phormica}. For example, magnetic-field gradients can encode navigation pathways, acoustic fields can create dynamically reconfigurable trapping landscapes, and optical fields can define spatially varying activity profiles\cite{nelson2010microrobots,ding2012chip,lozano2016phototaxis}. Individual particles need only respond to local field cues, while global coordination emerges from the structure of the field\cite{wu2021ion}. In this sense, the field acts not only as a control input but also as an external computational resource, closely related to concepts of morphological computation and embodied intelligence\cite{pfeifer2009morphological,auerbach2012relationship}. 

When field information is combined with feedback and adaptive control, collective information processing can emerge at the system level. Dynamic fields coupled to sensing and control loops allow active matter systems to continuously acquire information, update their behavior, and adapt to changing environments\cite{wang2022robots}. Examples include light-guided robot collectives that autonomously navigate complex terrains and particle robotic systems that achieve robust collective transport through distributed physical interactions\cite{lavergne2019group,reinhardt2025artificial,li2019particle}. For example, as shown in Fig. \ref{infocarrier}(c), in the Robo-Matter system integrated with an interactive feedback light-field environment, dynamic light spots carrying local activation information can selectively activate and follow individual robots according to their local physical environment and cluster structure, enabling onboard-computation-free collectives to adaptively regulate their collective structures and behaviors for tasks such as smart morphing and infiltration\cite{wang2024robo}. In such systems, information is no longer merely transmitted through the field; it becomes an active dynamical variable that shapes collective behavior. This perspective naturally bridges field-driven active matter with broader concepts of embodied intelligence, distributed computation, and information-powered nonequilibrium systems.

From this perspective, fields are not merely external control inputs but multifunctional physical infrastructures that simultaneously inject energy, mediate interactions, and encode information. The integration of these three roles provides a unifying framework for understanding field-driven active matter as an embodied physics–information system. This conceptual framework lays the foundation for the following sections, where robotic swarms are discussed as physical realizations of field-mediated physics–information coupling and as experimental platforms for exploring programmable nonequilibrium matter.
 
\section{\label{sec:scales}Robotic Swarms: Realizations of Field-Mediated Coupling Between Physics and Information}

\begin{figure}[t]
    \centering
\includegraphics[width=0.95\linewidth]{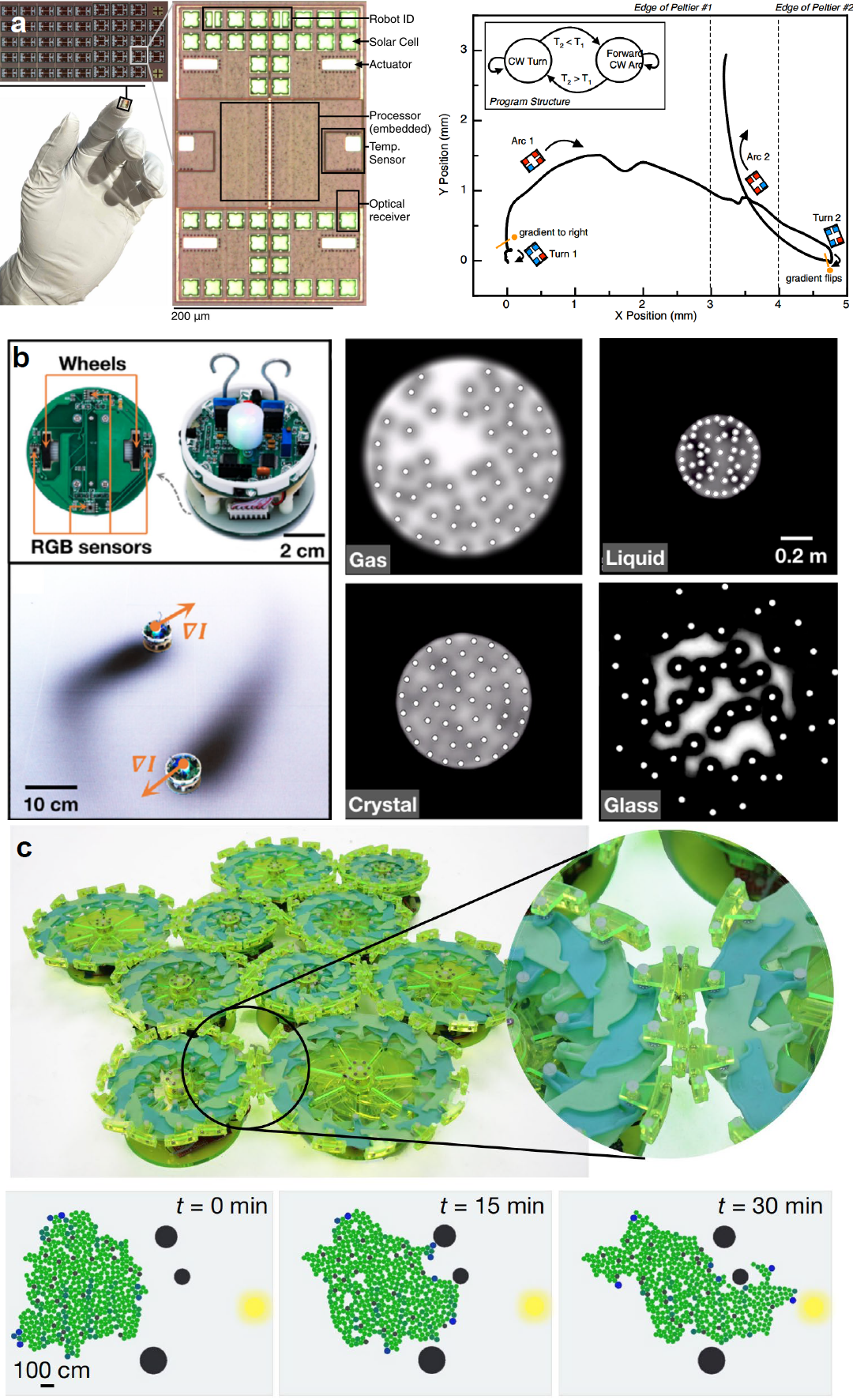}
\caption{\textbf{Robotic swarms provide experimentally accessible platforms in which energy, information, and collective dynamics become deeply intertwined.} (a) Based on fully lithographic processing, microrobots comparable in size to a single-celled paramecium can sense and compute, and therefore can climb temperature field gradients\cite{lassiter2025microscopic}. Reproduced with permission from Lassiter \textit{et al.}, Sci. Robot. \textbf{10}(109), eadu8009 (2025). Copyright 2025 The American Association for the Advancement of Science. (b) Wheeled Robots using RGB color sensors can interact with the light resource field to emerge dynamic and spatial transitions\cite{wang2021emergent}. Reproduced with permission from Wang \textit{et al.}, Phys. Rev. Lett. \textbf{126}(10), 108002 (2021). Copyright 2021 American Physical Society. (c) Particle robots can loosely couple with each other using dangling magnets and oscillate along own radius, able to move under the guidance of the light field based on statistical mechanics\cite{li2019particle}. Reproduced with permission from Li \textit{et al.}, Nature \textbf{567}(7748), 361-365 (2019). Copyright 2019 Springer Nature.}
\label{robotswarm}
\end{figure}

The development of robotic swarms across multiple length scales has transformed the traditional separation between mechanical actuation and information processing\cite{dorigo2021swarm}. From microscale biohybrid systems to macroscale self-reconfigurable collectives, robotic swarms provide experimentally accessible platforms in which energy injection, information processing, and collective dynamics become deeply intertwined\cite{zarepour2024biohybrid,ozkan2021self,lassiter2025microscopic,wang2021emergent,li2019particle}, as shown in Fig. \ref{robotswarm}. Through interactions mediated by external fields and environmental feedback, robotic bodies actively participate in sensing, information transduction, and collective decision-making\cite{wang2024robo}. As a result, robotic swarms constitute one of the most versatile physical realizations of field-mediated physics–information coupling and establish an important bridge between active matter theory and programmable intelligent materials\cite{mcevoy2015materials}. Accordingly, this section is organized around three complementary dimensions: microscale field-driven robotic swarms, macroscale robophysical systems, and embodied computational collectives.\\

\subsection{\label{subsec:Microscale}Microscale Robotic Swarms: From Propulsion to Control and Biohybrid Integration}

At microscopic scales, viscous forces dominate inertial effects, making locomotion fundamentally constrained by the kinematic reversibility of low-Reynolds-number hydrodynamics\cite{purcell2014life}. Consequently, microscale robotic swarms rely heavily on external physical fields to generate propulsion and coordinate collective motion. Magnetic, acoustic, optical, and electric fields provide remotely controllable energy inputs that allow large populations of micro- and nanorobots to move, assemble, and reorganize without requiring onboard power sources\cite{palagi2018bioinspired,chen2024materials}. Among the most widely studied examples are magnetically and acoustically driven microrobotic systems developed by Metin Sitti \textit{et al.}\cite{diller2011control,aghakhani2022high,mahkam2025multiorifice}. By programming spatially varying magnetic fields or acoustic pressure landscapes, these systems can guide the collective transport and reconfiguration of large populations of microscopic agents. Similarly, Peer Fischer \textit{et al.} developed helical magnetic microdrillers that exploit rotation–translation coupling to achieve nonreciprocal propulsion in highly viscous and viscoelastic biological environments\cite{ghosh2009controlled,schamel2014nanopropellers}. Such systems demonstrate how external fields can overcome the physical constraints of microscale locomotion and provide precise control over collective behavior.

Beyond propulsion, modern field-driven microswarms increasingly employ hybrid actuation schemes that combine multiple physical modalities\cite{chen2019hybrid,gao2011hybrid}. Acoustic–magnetic, optical–electrical, and other hybrid control architectures offer additional degrees of freedom for manipulating individual agents and collective configurations\cite{ahmed2021bioinspired,wang2025advanced,chiou2005massively,sun2023bioinspired}. Through dynamic modulation of external fields, microswarms can perform cargo capture, transport, release, and cooperative manipulation tasks with increasing precision\cite{yan2017multifunctional}. On the other hand, as shown in Fig. \ref{robotswarm}(a), advances in robotic microfabrication have recently enabled submillimeter microrobots with onboard sensing, programmable computation, and feedback, allowing autonomous field-responsive behaviors such as temperature-gradient climbing\cite{lassiter2025microscopic}. This complements field-driven microswarms by embedding part of the control intelligence directly on the microrobots themselves.

A further step toward functional intelligence emerges through the integration of living biological components with artificial robotic structures\cite{gwisai2022magnetic,mirkhani2024spatially}. Biohybrid microrobots combine the autonomous sensing, adaptation, and targeting capabilities of living cells with the controllability of engineered materials\cite{zhang2021dual,jin2023swarming,wang2023magnetically}. Beyond physical control, such systems increasingly incorporate functional biological autonomy for task-specific biomedical applications\cite{quan2026biophysics}. Recent studies have demonstrated biohybrid swarms composed of magnetically functionalized bacteria or stem-cell-based microrobots capable of collective navigation in physiological environments under programmable magnetic actuation\cite{wang2021magnetic,felfoul2016magneto,jeon2019magnetically}. Guided by rotating magnetic field combined with magnetic gradients, these cell-based robotic collectives can actively navigate through pulsatile biological fluids and accumulate at deep tissue targets. More broadly, such systems illustrate how biological functionality and field-mediated control can be integrated into a unified platform, blurring the boundary between living and artificial active matter.

\subsection{\label{subsec:Macroscale}Macroscale Robotic Collectives: From Discovering Physics to Engineering Function}

As system size increases from micrometers to centimeters, the dominant physical mechanisms shift from viscous hydrodynamics to contact mechanics, frictional interactions, and inertial effects. This transition has given rise to the field of robophysics, pioneered by Daniel I. Goldman \textit{et al.}, which uses simplified robotic systems as experimental physical models for exploring the principles governing collective behavior in complex, nonequilibrium environments\cite{aguilar2016review}. These studies typically use robots dominated by mechanical interactions, which rely on simple differential-driven or vibration-induced locomotion and local physical interactions rather than explicit communication\cite{li2022field,li2021programming,dmitriev2025swarmodroid}. By systematically varying robot geometry, interaction rules, and environmental conditions, these platforms revealed how collective organization can emerge directly from embodied short-range interactions. For example, as shown in Fig. \ref{robotswarm}(b), a swarm of field-driven robots, which move only in response to self-generated light-field gradients created by their own resource consumption, exhibits density-dependent transitions from gas to crystal, liquid, and jammed states, illustrating resource-mediated collective emergent behaviors\cite{wang2021emergent}. In this sense, robotic swarms served as controllable model systems for studying active matter phenomena in a manner analogous to how model organisms are used in biology\cite{savoie2019robot}. 

More recently, an increasing emphasis has gradually shifted from using robots to discover physical principles toward exploiting those principles to engineer functional materials. This transition is exemplified by the emergence of robotic matter systems capable of dynamically altering their collective mechanical properties. Representative examples include the Magbots (see Fig. \ref{infocarrier}(c)) and the Granulobot platform, in which individual units equipped with internal actuators generating symmetry-breaking motion interact through reversible magnetic contacts\cite{wang2024robo,saintyves2024self}. Under external driving and local perturbations, these units can self-organize into mechanically coherent structures while retaining the ability to flow and reconfigure when necessary. Such systems demonstrate how collective robotic behavior can be harnessed to create adaptive materials whose properties emerge from distributed physical interactions rather than centralized control\cite{sun2026recent,goldman2024robot}. The broader significance of these developments lies in the gradual convergence of robotics and materials science. Instead of viewing robots as discrete agents executing tasks independently, robotic collectives increasingly function as programmable matter whose macroscopic properties arise from the coordinated behavior of many interacting units\cite{gilpin2010modular,chafik2024conventional}. This shift establishes robotic swarms as a promising platform for realizing adaptive, multifunctional materials with dynamically tunable mechanical and transport characteristics.

\subsection{\label{subsec:PassToEmbodi}From Passive Actuators to Embodied Computational Entities}

Perhaps the most profound conceptual shift enabled by robotic active matter is the transition from viewing robots as passive actuators to treating them as embodied computational entities. In this emerging paradigm, information processing is no longer necessarily confined to centralized processors or explicit communication networks. Instead, physical morphology, environmental interactions, and collective dynamics themselves participate directly in computation\cite{pfeifer2007self,paul2006morphological,fuchslin2013morphological}. Rather than treating the body as a passive executor of control commands, embodied computation exploits the dynamics of the body and its interaction with the environment to simplify sensing, decision-making, and control\cite{rus2015design,hauser2012role,liu2025embodied}. As a result, information processing becomes distributed across both physical and informational degrees of freedom\cite{devlin2025material}.

\begin{table*}[!ht]
\caption{\label{tab:field_driven_platforms}
Representative field-driven experimental platforms mapped across the complementary dimensions of platform scale, interaction transmission mechanism, field modality and role, collective dynamics, and functionality.}
\begin{ruledtabular}
\begin{tabular}{p{2.5cm} p{1.5cm} p{2.0cm} p{1.8cm} p{3.2cm} p{3.1cm} p{2.7cm}}
Representative system & 
Platform scale & 
Interaction mechanism$^{1}$ & 
Field modality & 
Dominant field role(s) $^{2}$ & 
Collective dynamics & 
Function / material property
\\
\hline

Resource-field-driven robot swarm ~\cite{wang2021emergent}
&
Macroscale
&
Field-mediated
&
Optical
&
Interaction mediator; information carrier
&
Field-controlled collective states and phase-like transitions
&
Adaptive collective behavior
\\

Robo-Matter / Magbots~\cite{wang2024robo}
&
Macroscale
&
On-contact; short-range; information-mediated
&
Optical
&
Interaction mediator; information carrier
&
Dynamic self-assembly, morphing, infiltration, and self-reorganization
&
Reconfiguration; self-repair
\\

Hematite colloidal microrobots ~\cite{xie2019reconfigurable}
&
Microscale
&
On-contact; short-range; field-mediated
&
Magnetic; hydrodynamic
&
Energy injection; interaction mediator; information carrier
&
Reconfigurable collective morphologies and coordinated locomotion
&
Structural reconfiguration; cargo manipulation
\\

Particle robots~\cite{li2019particle}
&
Macroscale
&
On-contact; short-range
&
Optical
&
Information carrier
&
Mechanically coupled collective waves and locomotion
&
Embodied computation; collective locomotion
\\

Anatase TiO$_{2}$ colloidal nanomotors~\cite{choudhury2019chemical}
&
Microscale
&
Field-mediated
&
Optical; hydrodynamic
&
Energy injection; interaction mediator
&
Dense active self-organization and activity-dependent rheology
&
Tunable mechanical and rheological properties
\\

Fe-Ni alloy microrobot collectives~\cite{xu2025reconfigurable}
&
Microscale
&
Field-mediated
&
Magnetic
&
Energy injection; interaction mediator
&
Robust collective assembly and force amplification
&
Collective force generation; mechanical manipulation
\\

Ferromagnetic Ni spheres~\cite{han2024self}
&
Microscale
&
Short-range; field-mediated
&
Magnetic; hydrodynamic
&
Energy injection; interaction mediator
&
Self-assembled pump architectures and collective flow
&
Active pumping and cargo transport
\\

Superparamagnetic microsphere microrobot swarms~\cite{wang2026artificial}
&
Microscale
&
On-contact; short-range; field-mediated
&
Magnetic; hydrodynamic
&
Energy injection; interaction mediator; information carrier
&
Swarm unraveling and collective tangential flow
&
Directed cargo transport under confinement
\\

Paramagnetic colloidal carpets~\cite{massana2019tunable}
&
Microscale
&
Short-range; field-mediated
&
Magnetic; hydrodynamic
&
Energy injection; interaction mediator
&
Field-controlled treadmilling and structural regeneration
&
Collective self-repair
\\

Paramagnetic Fe$_{3}$O$_{4}$ nanoparticles ~\cite{jin2019mimicking}
&
Microscale
&
On-contact; short-range; field-mediated
&
Magnetic; hydrodynamic
&
Energy injection; interaction mediator; information carrier
&
Field-driven assembly and reversible structural elongation
&
Self-repair; functional restoration
\\

Polymer microbead reservoir computer~\cite{wang2024harnessing}
&
Microscale
&
Field-mediated; information-mediated
&
Optical
&
Energy injection; information carrier
&
Delayed nonlinear dynamics and fading memory
&
Physical information processing
\\

\end{tabular}
\vspace{2pt} 
\begin{minipage}{\textwidth} 
\footnotesize
\noindent
\raggedright
$^{1}$ Interaction mechanism describes how interactions propagate through the system, as discussed in Sec.~\ref{sec:need}. 

\noindent
\raggedright
$^{2}$ Field role specifies the primary physical or informational function of the field, as discussed in Sec.~\ref{sec:fields}.
\end{minipage}
\end{ruledtabular}
\end{table*}

A representative realization of this idea is the Particle Robotics framework developed by Hod Lipson \textit{et al.}\cite{li2019particle}, as shown in Fig. \ref{robotswarm}(c). From the perspective of embodied computation, particle robots perform a sensorimotor transformation that converts environmental field information into collective motion. Each particle robot locally senses the environmental light-field signal and encodes its intensity into the phase of its radial expansion–contraction cycle. Mechanical coupling among particle robots then integrates these local phase differences into spatiotemporal waves, whose interaction with the substrate generates directed motion of the collective based on statistical mechanics. Thus, sensing, information integration, and motor control are partially embedded in the mechanical dynamics themselves, with phototaxis emerging directly from the physical transformation of a spatial light-field gradient into coordinated locomotion. In other words, in these systems, large populations of simple robotic particles possess extremely limited individual capabilities and rely on global actuation signals rather than local communication or centralized planning. Nevertheless, collective locomotion and navigation emerge through the statistical organization of physical interactions among particles. Such examples demonstrate how reliable collective behavior can arise from the interplay between mechanical constraints, stochastic fluctuations, and field-mediated coordination. Related ideas have also been explored in robotic systems composed of shape-changing active particles, such as smarticles\cite{savoie2019robot}. In these platforms, coordinated collective behaviors emerge through repeated mechanical interactions among agents, without requiring explicit communication channels. Physical collisions, deformation dynamics, and state-dependent interactions can effectively realize information-processing functions that would traditionally be implemented algorithmically. Consequently, sensing, computation, and control become distributed throughout the collective and partially embodied within the material substrate itself.

From a broader perspective, embodied computation provides a conceptual bridge between robotic swarms and information-processing materials\cite{ben2023morphological}. By embedding aspects of sensing, memory, and decision-making directly into physical interactions, robotic active matter moves beyond the conventional paradigm of robots as isolated machines\cite{sharma2020morphological}. Instead, it points toward a new class of intelligent materials in which information processing emerges from the collective dynamics of matter itself\cite{kaspar2021rise,miskin2026robotic}. Across scales, robotic swarms progressively transform external fields from simple energy sources into channels for interaction mediation and information processing. In doing so, they provide experimentally accessible realizations of the physics–information coupling framework introduced in Section \ref{sec:fields}. Rather than serving merely as engineered systems, robotic swarms increasingly function as programmable active matter whose collective behavior emerges from the co-design of energy injection, interaction topology, and information flow. This perspective naturally leads to the nonequilibrium phenomena discussed in the following section. 

To facilitate comparison across these complementary perspectives, Table~\ref{tab:field_driven_platforms} summarizes representative experimental platforms discussed throughout this review according to their physical scale, interaction-transmission mechanism, field modality and role, collective dynamics, and demonstrated functionality. The table is intended not as a hierarchical classification, but as a cross-sectional map linking the different dimensions introduced in this review. A given platform may therefore appear in multiple categories, reflecting the fact that interaction mechanism, field role, and function constitute complementary descriptors of the same underlying field–agent–environment system.

\section{\label{sec:nonequi}Nonequilibrium Physical Phenomena in Field-Driven Physics-Information Coupled Systems}

\subsection{\label{subsec:Classical}Classical Nonequilibrium Phenomena in Active Matter: Self-Propulsion and Emergent Collective Behavior}

A defining characteristic of classical active matter is that individual agents continuously consume locally available energy sources, such as chemical or electrical energy, and convert them into directed mechanical motion, thereby maintaining the system in a persistent nonequilibrium state\cite{narayan2007long}. When self-propelled particles interact through purely repulsive collisions, their effective propulsion speed often decreases with increasing local density. Within certain parameter regimes, this velocity–density negative feedback can give rise to motility-induced phase separation (MIPS), causing the system to spontaneously segregate into dense condensed phases and dilute gaseous phases\cite{gonnella2015motility}. Statistical-physics studies have revealed that this self-trapping mechanism is associated with a characteristic critical point. However, the critical behavior of MIPS remains an active topic of debate. Some studies suggest that its long-wavelength critical properties are closely related to those of the equilibrium Ising universality class\cite{partridge2019critical}, whereas other theoretical and numerical investigations report significant deviations\cite{dittrich2021critical}, reflecting the influence of nonequilibrium driving on critical fluctuations. In contrast, polar active fluids, such as bird flocks or fish schools, exhibit spontaneous collective motion through alignment interactions, leading to the emergence of large-scale flocking states\cite{vicsek1995novel,kudrolli2008swarming}. Importantly, MIPS and flocking represent fundamentally different classes of nonequilibrium phase transitions. MIPS originates from the density-dependent slowdown of scalar active particles and does not require orientational order, whereas flocking arises from active self-propulsion combined with symmetry-breaking orientational interactions and supports long-range order even in two dimensions. Thus, unlike equilibrium systems constrained by the Mermin–Wagner theorem, polar flocking systems are not subject to these restrictions due to their intrinsically nonequilibrium dynamics and associated advective nonlinearities\cite{toner1995long}. Furthermore, under dense confinement, strong nonlinear networks of mechanical contacts can induce jamming transitions, leading to dynamically arrested states\cite{liu2010jamming}. Robotic active-matter platforms have provided experimentally accessible realizations of many of these phenomena\cite{scholz2018rotating}. Examples include MIPS-like clustering in BOBbots, entanglement-driven collective states in robotic worm blobs, and jamming-like transitions in dense robotic assemblies\cite{li2021programming,adorjani2024motility,ozkan2021collective,fitzgerald2020review}. These systems demonstrate that the underlying mechanisms of active-matter self-organization are remarkably robust across physical scales.

\subsection{\label{subsec:UniquetoFieldDriven} Field-Induced Nonequilibrium Phenomena: Nonreciprocity and Programmable Criticality}

External physical fields significantly expand the range of nonequilibrium phenomena accessible to active matter\cite{yu2023programmable}. Beyond serving as energy sources, magnetic, electric, acoustic, and optical fields dynamically reshape effective interactions among active agents, introducing coupling mechanisms that are not typically accessible or easily tunable in conventional self-propelled systems\cite{hou2023review}. As a result, collective behavior is no longer determined solely by particle-level activity and local collisions, but also by externally programmable interaction networks that can be reconfigured in space and time\cite{wang2023selective}. A central consequence of field mediation is the emergence of effectively long-range and dynamically tunable interactions. In magnetically driven systems, field-induced dipolar forces generate anisotropic attractions that promote chain formation, clustering, and collective synchronization\cite{ceron2023programmable,driscoll2017unstable}. In rotating colloidal or microrobotic assemblies, hydrodynamic flows generated by particle motion introduce additional coupling channels whose strength depends on driving frequency, fluid properties, and particle geometry\cite{martinez2018emergent}. The competition between these interaction mechanisms can lead to collective-state transitions that are inaccessible in systems governed only by short-range steric interactions\cite{wang2021emergent,liu2023orderly}. More generally, external fields transform interactions from static constraints into actively controllable dynamical variables.

\begin{figure}[!t]
    \centering
\includegraphics[width=0.95\linewidth]{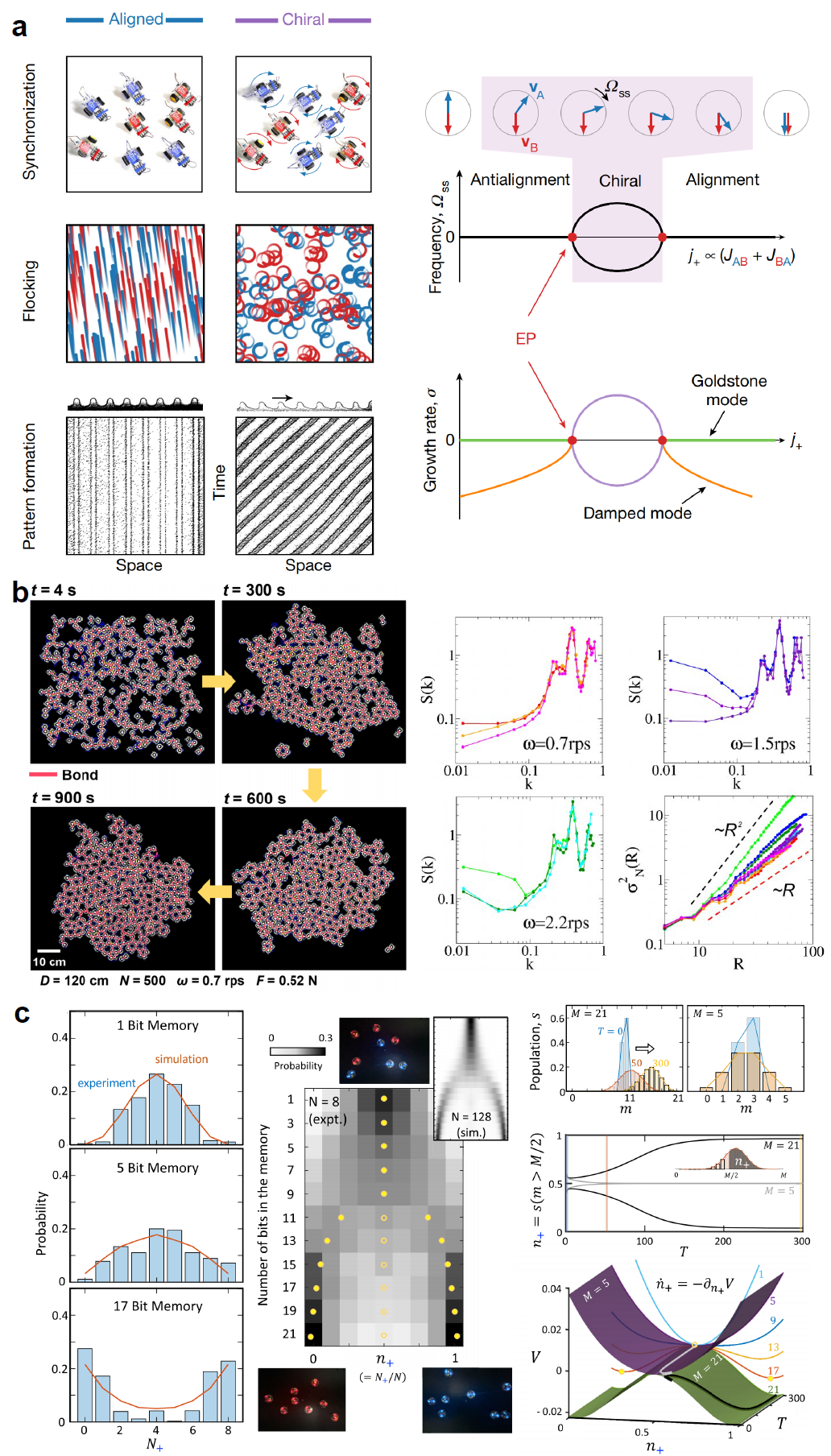}
\caption{\textbf{An increasingly rich hierarchy of nonequilibrium phenomena emerges in physics-information coupled robotic active matter.} (a) The phase transition from static alignment to a chiral motion that spontaneously breaks parity emerges in robots programmed as non-reciprocal spins\cite{fruchart2021non}. Reproduced with permission from Fruchart \textit{et al.}, Nature \textbf{592}(7854), 363-369 (2021). Copyright 2021 Springer Nature. (b) Originating from the competition between tunable binding and local active twist, Magbots driven by the light field can achieve disordered hyperuniform structures\cite{wang2025hyperuniform}. Reproduced with permission from Wang \textit{et al.}, Phys. Rev. Lett. \textbf{134}(24), 248301 (2025). Copyright 2025 American Physical Society. (c) Memory-induced spontaneous symmetry breaking emerges in an air-table drone system endowed with internal memory and a binary decision variable\cite{li2026informational}. Reproduced with permission from Li \textit{et al.}, Phys. Rev. Lett. \textbf{136}(13), 138302 (2026). Copyright 2026 American Physical Society.}
\label{nonequilibrium}
\end{figure}

Field-mediated interactions also provide a natural route toward nonreciprocity. In equilibrium systems, effective pair interactions are typically reciprocal and obey action–reaction symmetry. By contrast, when particle dynamics are coupled to externally imposed or self-generated fields, effective interactions may become asymmetric, such that the influence of particle A on particle B differs from that of particle B on particle A\cite{fruchart2021non,soares2026dissipative}, as shown in Fig.\ref{nonequilibrium}(a). Nonreciprocal interactions have been observed in chemically active colloids, hydrodynamically coupled driven rotors, and some field-driven chiral active matter systems\cite{wu2021ion,wang2015one,liebchen2022interactions}. Such asymmetry fundamentally alters collective dynamics, enabling behaviors including persistent chasing, self-organized oscillations, and dynamically stabilized nonequilibrium structures that have no equilibrium counterparts\cite{ji2023collective,chao2026selective}. 

At continuum scales, nonreciprocal interactions can manifest as emergent transport and mechanical properties that violate conventional equilibrium constraints\cite{brandenbourger2019non,veenstra2024non}. Particularly notable examples include odd viscosity and odd elasticity, which arise in driven systems that simultaneously break time-reversal and parity symmetries\cite{chen2021realization,scheibner2020odd,fruchart2023odd}. Unlike conventional viscosity and elasticity, these responses contain antisymmetric components that permit energy transfer between orthogonal modes of deformation. Consequently, the constitutive relations governing stress and strain are no longer constrained by equilibrium reciprocity, allowing the material to perform nonzero mechanical work during closed deformation cycles. Chiral active-spinner systems have emerged as prototypical experimental platforms for investigating these unusual forms of nonequilibrium mechanics and have revealed a growing class of active materials whose behavior cannot be fully described within traditional hydrodynamic or elastic frameworks\cite{yang2021topologically,banerjee2017odd}.

The influence of external fields extends beyond transport properties to the statistical physics of collective-state transitions. By dynamically modifying interaction networks and fluctuation spectra, external fields can reshape phase-transition pathways and modify critical behavior\cite{wang2021emergent,bechinger2020active}. Field-controlled active systems have been shown to suppress Ostwald ripening during phase separation, stabilize microphase-separated states, and generate long-lived nonequilibrium patterns\cite{zwicker2015suppression,bressloff2020active,tjhung2018cluster,huang2021circular}.  Recent studies of magnetically coupled active spinner networks have further demonstrated how external fields can be used to engineer hyperuniform states and tune collective fluctuations over large length scales\cite{wang2025hyperuniform,ma2025simple,lei2019nonequilibrium}, as shown in Fig.\ref{nonequilibrium}(b). Together, these results illustrate how field mediation transforms phase behavior from a passive consequence of microscopic interactions into an actively programmable property of nonequilibrium matter.

\subsection{\label{subsec:InformationCoupled} Information-Coupled Nonequilibrium Phenomena: Feedback, Memory, and Information-Driven Ordering}

The incorporation of sensing, memory, and decision-making introduces an additional layer of dynamical feedback beyond those arising from energy consumption and physical interactions alone. In such systems, information is no longer merely a descriptor of the system state but becomes an active dynamical resource capable of influencing collective behavior. Unlike energy injection or interaction modification, information processing introduces additional internal state variables associated with sensing, memory, prediction, and decision-making\cite{li2026informational}. These variables influence system evolution without being reducible to conventional mechanical interactions alone. The thermodynamic significance of information has long been recognized through concepts such as Maxwell's demon, Landauer's principle, and information thermodynamics\cite{landauer1961irreversibility,bennett1982thermodynamics,parrondo2015thermodynamics}. In active systems equipped with feedback mechanisms, individual agents can measure local environmental conditions and adapt their behavior accordingly\cite{sagawa2013information}. Such sensing–decision loops enable acquired information to be converted into directed work through feedback control, effectively biasing stochastic dynamics toward desired outcomes. However, this enhanced control is not obtained for free. Information acquisition, storage, and especially information erasure require energy expenditure, and generalized fluctuation theorems establish rigorous quantitative relationships between information flow and entropy production\cite{sagawa2010generalized,sagawa2012thermodynamics,ro2022model}. From this perspective, collective ordering can be understood as a process in which information continuously constrains dynamical degrees of freedom, leading to an effective reduction of accessible phase space at a finite thermodynamic cost.

Information feedback can also substantially alter collective-state transitions. In contrast to conventional active matter, where phase behavior is determined primarily by interaction strengths and activity levels, feedback-controlled systems introduce additional control parameters associated with feedback gain, sensing accuracy, communication fidelity, and delay time\cite{attanasi2014information,cavagna2013boundary,roy2019effect,pakpour2024delay}. Theoretical studies have shown that the amount of accessible information can directly influence collective transport and coordinated behavior, producing transitions that are regulated not only by mechanics but also by information-processing capabilities\cite{davis2024active,cappetto2025digitization}. Such information-driven ordering phenomena suggest that collective-state selection may depend not only on physical interactions but also on information flow. An additional consequence of information coupling is the emergence of memory-dependent collective behavior. Most classical active-matter models assume that particle dynamics depend only on the current state of the system. However, many biological and engineered agents retain information about past interactions, introducing temporal correlations into their dynamics\cite{na2021bio,salman2020phormica}. Memory can generate effective long-range interactions in time, alter stability boundaries, and create entirely new collective states\cite{bechinger2016active}. Recent theoretical and experimental studies have demonstrated that delayed feedback, finite-memory kernels, and history-dependent interactions can induce oscillatory phases, bistable states, and memory-stabilized collective order\cite{li2026informational,horton2025order}. These findings suggest that memory should be regarded as an independent dynamical variable capable of reshaping nonequilibrium phase behavior.

Recent robotic active-matter platforms provide experimentally accessible realizations of these concepts. The Kilobot system demonstrates how collective locomotion can emerge from embodied information processing despite extremely limited individual capabilities\cite{rubenstein2014programmable,slavkov2018morphogenesis}. More recently, memory-enabled robotic systems have shown that collision histories encoded within finite memory windows can bias local decision-making and drive collective symmetry breaking\cite{li2024memory,li2026informational}, as shown in Fig.\ref{nonequilibrium}(c). Such systems illustrate how information storage, feedback, and memory can interact with physical dynamics to influence macroscopic collective behavior. More broadly, they provide a bridge between information thermodynamics and active matter, revealing how information-processing mechanisms can become integral components of nonequilibrium self-organization.

In this sense, \textit{field-driven robotic active matter} can be seen as a new class of active systems in which externally imposed or self-generated fields inject energy, mediate interactions, and shape collective dynamics, while onboard or distributed information processing allows the robots to respond adaptively to those fields.  Key features of these systems are that their dynamics can be programmed, their interactions can be mediated through physical fields, and their collective behavior can change in response to environmental information. For example, robots moving on a deformable elastic membrane may interact through the membrane deformation they collectively produce: each robot changes the local surface profile, and this deformation then biases the motion of other robots. Another example is a group of robots moving on a shared resource landscape \cite{wang2021emergent}, where the robots consume, deplete, or redistribute a field that affects their future motion (see Fig. \ref{robotswarm}(b)). In these cases, the interaction is not simply pairwise or instantaneous. Instead, it is mediated by an extended environmental field that can store, transmit, and transform information.

A minimal theoretical example is emergent bootstrapped motion on an adaptive resource landscape \cite{phan2021bootstrapped}. In that model, an agent consumes or modifies a field, and the modified field then feeds back on the agent's future motion. The agent therefore does not move only in response to its present state; it also responds to a landscape that has been shaped by its own past trajectory. In this sense, the field becomes an \textit{active memory medium}: it stores information about where the agent has been, how strongly it interacted with the environment, and how the environment has recovered or evolved afterward. That stored information is then converted into a future force, bias, or effective rule of motion.

This model is relatively simple, but it captures the essential point of field-driven active matter: nontrivial motion can emerge once the field is dynamical rather than a fixed background. The agent writes information into the field, and the field converts that stored information back into motion. For this reason, we briefly sketch the mathematics below, not as a complete theory, but as the simplest prototype of field-driven dynamics-information coupling. 

\begin{figure}[h]
    \centering
\includegraphics[width=\linewidth]{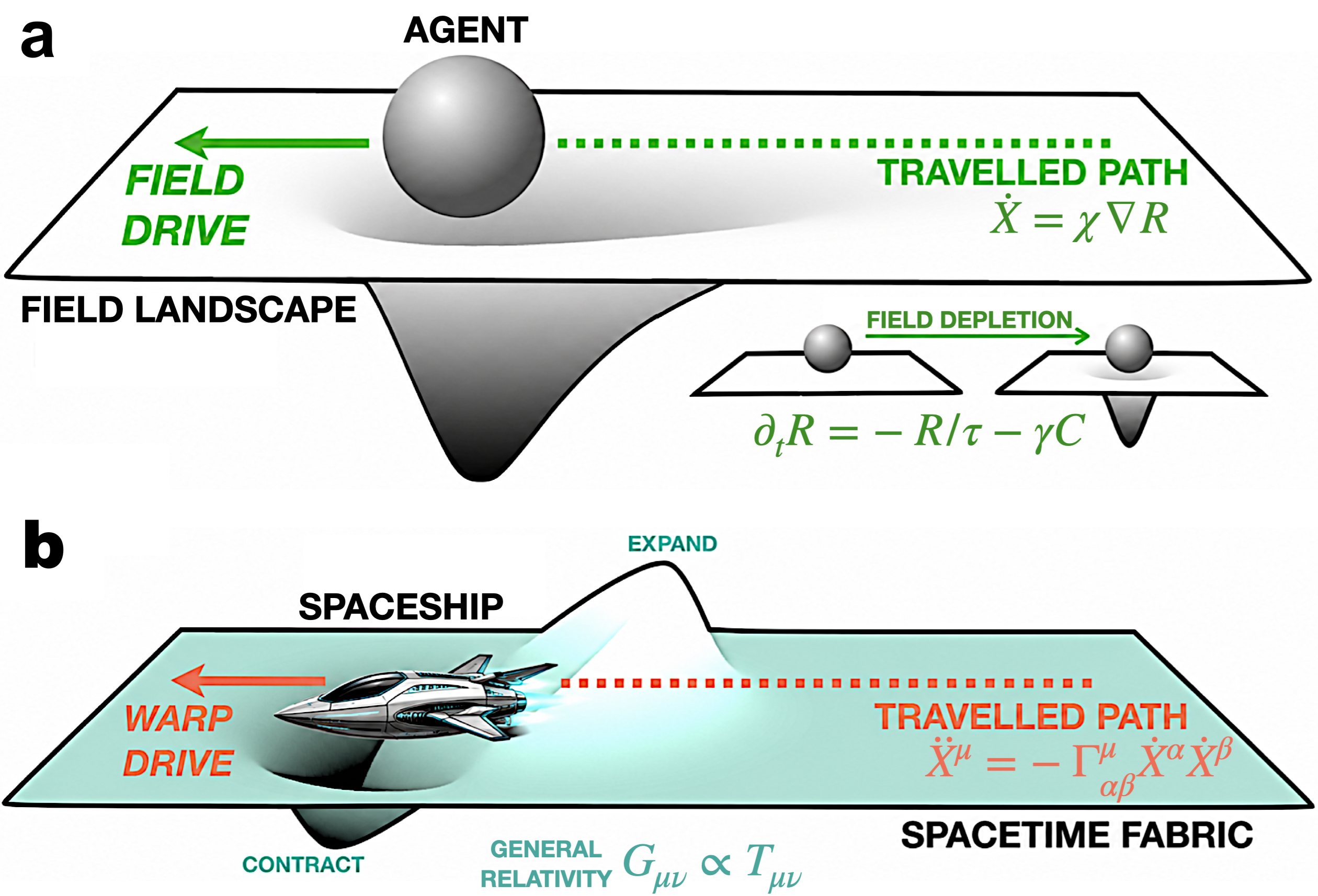}
    \caption{\textbf{Field-driven analogy.} Bootstrapped motion can emerge when an agent modifies its surrounding background, and the resulting background reaction feeds back on the agent's motion. Reproduced with permission from Wang \textit{et al.}, Phys. Rev. Lett. \textbf{126}(10), 108002 (2021). Copyright 2021 American Physical Society. (a) Field driven motion\cite{wang2021emergent}. (b) Warp drive analogy in General Relativity \cite{alcubierre1994warp}.}
\label{fig01}
\end{figure} 

Consider a one-dimensional space. Let $R(x,t)$ denote a recovering field, and let $X(t)$ be the position of an agent. We assume that the agent moves according to a local sensor measurement of the field gradient:
\begin{equation}
    \frac{d}{dt} X(t) = \chi \, \partial_x R(x,t)\Big|_{x=X(t)} \ ,
    \label{eq:agent_motion}
\end{equation}
where $\chi$ characterizes the strength of the agent's response to the field gradient. This is a robotic analogue of bacterial chemotaxis that gives rise to collective transport behavior \cite{nguyen2024remark,phan2024direct}. The field itself evolves according to a recovering dynamics with relaxation time $\tau$, meaning it decays back to zero in the absence of forcing as
\begin{equation}
    \partial_t R(x,t)
    =-R(x,t)/\tau \ .
\end{equation}
In addition, it is locally depleted by a moving agent at position X(t), giving
\begin{equation}
    \partial_t R(x,t)
    =- R(x,t)/\tau - \gamma C(r^2)\Big|_{r=\frac{x-X(t)}{\lambda}} \ .
    \label{eq:field}
\end{equation}
Here $C$ is a normalized consumption profile with the peak value $C(0)=1$, $\gamma$ controls the depletion strength, and $\lambda$ sets the spatial range of the influence.

If the field were fixed at the equilibrium value $R(x,t)=0$, Eq. \eqref{eq:agent_motion} would simply describe a random walk motion. But here the landscape is not prescribed but dynamically generated by the agent itself as in Eq. \eqref{eq:field}. To see how self-sustained motion can emerge, we look for a constant-velocity solution of the form $$x=Wt \ . $$ We first solve for the corresponding field $R(x,t)$ generated by this steadily moving agent using the field dynamics Eq. \eqref{eq:field}, and then substitute it back into the agent response Eq. \eqref{eq:agent_motion} to obtain a self-consistency \textit{bootstrap} condition for the velocity as follows:
\begin{equation}
\begin{split}
    W &= - \frac{2\chi \gamma}{\lambda^2} \int^{+\infty}_0 dt \ Wt e^{-t/\tau} C'(r^2) \Big|_{r=\frac{Wt}{\lambda}} \ 
    \\
    &\approx \frac{2 \chi \gamma |C'(0)|}{(\lambda/\tau)^2} W - \frac{12 \chi \gamma C''(0)}{(\lambda/\tau)^4} W^3 + \mathcal{O}(W^5) \ ,
\end{split}
\label{eq:get_W}
\end{equation}
where the second line follows from expanding the right-hand side in powers of small $W \rightarrow 0$, as appropriate near the \textit{onset} of \textit{bootstrapped motion}. Eq. \eqref{eq:get_W} always admits the trivial solution $W=0$. However, a nonzero solution $W\neq 0$ appears only when the feedback between the agent and the field is sufficiently strong, i.e.
\begin{equation}
    \frac{2 \chi \gamma |C'(0)|}{(\lambda/\tau)^2} > 1 \ .
\end{equation}
For example, the gradient following coefficient $\chi$ and the consumption rate $\gamma$ have to be large enough so that their product exceeds a critical value:
$$\chi \gamma > (\chi \gamma)_c \equiv \frac{(\lambda/\tau)^2}{2|C'(0)|} \ . $$
This condition has a simple physical interpretation: $\gamma$ controls how strongly the agent modifies the field, while $\chi$ controls how strongly the agent responds to the resulting field gradient. Their product therefore measures the strength of the feedback loop between motion and field deformation. Bootstrapped motion appears only when this feedback is strong enough to overcome the relaxation rate $1/\tau$ and spatial smoothing length $\lambda$ of the field. 

The bootstrapped motion emerges through a \textit{pitchfork bifurcation}: below a critical constraint, the only stable state is the stationary state; above it, the system develops two symmetry-related moving states:
\begin{equation}
    W \propto \pm \sqrt{\frac{\chi\gamma}{(\chi \gamma)_c} - 1} \ .
\end{equation} 
The two signs correspond to spontaneous symmetry broken motion to the left or to the right. The agent alone has no intrinsic directional preference, but when the agent modifies the field and then responds to the field it has modified, the coupled system can generate a new dynamical state. In this sense, the field is not a passive background. It is a dynamical memory medium that enables self-generated, history-dependent motion.

This field-driven paradigm also suggests a useful analogy with general relativity (see Fig. \ref{fig01}). In Wheeler's famous phrase, matter tells spacetime how to curve, and spacetime tells matter how to move \cite{wheeler2000geons}. Similarly, in field-driven robotic active matter, the robots reshape the field, and the reshaped field tells the robots how to move. The analogy is not meant to imply that the robotic system is gravitational. Rather, it emphasizes the reciprocal structure of the coupling: the agents are not moving through a fixed background, but through a dynamical medium that they themselves modify. The environment is therefore not just a passive substrate. It becomes part of the active system.

This viewpoint naturally leads to what we call \textit{physics-information-coupling feedback}. The physical dynamics of the robots generate information in the field, while the information stored in the field feeds back into the physical dynamics. Memory and learning can then arise not only from internal computation onboard each robot, but also from the collective modification of the environment itself. Recent work on informational active matter emphasizes that information processing, memory, and feedback can fundamentally alter the behavior of active systems \cite{li2026informational}. Field-driven robotic active matter provides a concrete platform where these ideas can be implemented, measured, and programmed. The field supplies a physical channel for interaction, the robots supply sensing and actuation, and the feedback between them produces adaptive collective behavior. Thus, the central advance of the field-driven robotic active-matter paradigm is to combine three ingredients: energy injection through fields, information processing through robotic agents, and reciprocal feedback through a shared dynamical environment. This combination bridges nonequilibrium physics and functional materials. It allows active matter to move beyond spontaneous self-organization toward programmable, adaptive, and task-oriented collective behavior.

Taken together, these developments reveal an increasingly rich hierarchy of nonequilibrium phenomena. Classical active matter is governed primarily by energy consumption and local interactions, field-driven systems introduce dynamically programmable interaction networks and nonreciprocal couplings, while information-enabled systems further incorporate sensing, memory, and feedback as active dynamical ingredients. As a result, nonequilibrium behavior can no longer be understood solely through the traditional language of forces and transport. Instead, collective dynamics emerge from the intertwined evolution of energy flow, interaction structure, and information processing. This observation motivates the need for a unified theoretical framework capable of describing how physical and informational degrees of freedom coevolve in active matter systems.

\section{\label{sec:Coupling}Physics–Information Coupling: A Unified Framework}

The transition from classical active matter to field-driven active matter requires a conceptual framework in which information is treated not merely as an external control signal but as a physical variable that actively participates in nonequilibrium dynamics, i.e., a bidirectional coupling between information and dynamics. Within this perspective, information exists across multiple organizational levels. At the agent level, information is stored in internal states, memory variables, and decision rules, representing algorithmic information that directly biases local dynamics; at the collective level, it emerges through correlations, fluctuations, and order parameters, forming statistical information that characterizes the collective state of the system; and at the material level, it becomes embedded within spatial organization, topological structures, and collective configurations, giving rise to structural information that can persist even after the original driving conditions have changed. These different forms of information continuously transform into one another through collective dynamics. Consequently, information is no longer a passive descriptor of system behavior but an active component of nonequilibrium evolution. This viewpoint naturally leads to a unified physics--information framework based on two complementary mappings: Information-to-Physics and Physics-to-Information, which together form a closed feedback loop linking computation, collective dynamics, and material functionality.\\

\subsection{\label{subsec:InToPhy}Information-to-Physics: Information as a Dynamical Control Variable}
The first mapping describes how information shapes physical dynamics. In field-driven active matter, external fields generate high-dimensional information landscapes through spatiotemporal modulation of parameters such as amplitude, frequency, phase, and spatial gradients. Simultaneously, internal memory states, feedback policies, and decision rules provide additional informational constraints on individual and collective behavior. Rather than acting solely as energy sources, these informational inputs modify the dynamical rules governing system evolution.

From a statistical-physics perspective, information enters the system by altering transition probabilities, interaction rules, and the effective stochastic dynamics\cite{sagawa2012fluctuation,dabelow2019irreversibility}. As a result, information effectively biases the exploration of phase space and redirects the evolution of collective states\cite{fodor2022irreversibility}. In feedback-controlled active systems inspired by Maxwell's demon, for example, agents selectively modify their control actions or interaction rules based on locally acquired information, thereby converting environmental fluctuations into directed collective motion\cite{vansaders2026measurement,negi2024collective}. More generally, information-guided interactions can reshape collective-state transitions, alter ordering pathways, and generate nonequilibrium steady states that would be inaccessible through purely mechanical interactions alone\cite{takatori2025feedback}.

Memory further extends this mechanism by introducing temporal correlations into the dynamics. Instead of responding only to instantaneous environmental conditions, agents can utilize information accumulated from past experiences to influence future behavior\cite{shaebani2020computational}. Such history-dependent dynamics effectively introduce additional degrees of freedom into the system and enable adaptive responses to heterogeneous and time-varying environments\cite{wang2022robots}. Consequently, information functions as a dynamical control variable that continuously reshapes the physical evolution of active matter.

\subsection{\label{subsec:PhyToIn}Physics-to-Information: Collective Dynamics as Information Processing}
The reverse mapping describes how physical dynamics generate, propagate, and store information. In this perspective, field-driven active matter operates as a distributed physical information-processing system in which computation emerges directly from collective dynamics rather than from centralized processors.

Information propagation arises through correlations and collective modes\cite{toner2005hydrodynamics}. Near collective-state transitions, active systems often exhibit enhanced susceptibility and long-range correlations, allowing local perturbations to spread efficiently across the entire collective\cite{cavagna2010scale}. Such critical-like regimes can significantly enhance the transmission of information and increase the responsiveness of the system to environmental inputs. 

Information storage emerges through persistent physical structures and collective organization. Long-lived vortices, hyperuniform networks, topological defects, and other emergent configurations can retain signatures of previous interactions, external driving histories, or environmental constraints\cite{shankar2022topological,paulsen2025mechanical,mungan2025self}. In this sense, collective structures function as physical memory substrates that encode information directly within the material organization of the system. 

Information computation arises from the nonlinear dynamics of collective behavior. External inputs are transformed through many-body interactions into collective responses that perform filtering, amplification, classification, and decision-making functions\cite{nakajima2015information}. This concept is closely related to physical reservoir computing and morphological computation, where information processing is distributed across the physical dynamics of the system itself\cite{tanaka2019recent,nakajima2014exploiting,kobayashi2025information,wang2024harnessing}. Rather than separating computation from embodiment, field-driven active matter can perform computation through the evolution of its physical dynamics.

\subsection{\label{subsec:CLoop}Closed-Loop Co-Evolution of Physics and Information}
The two mappings described above are not independent processes but are continuously coupled through feedback. Information modifies physical dynamics, while physical dynamics generate new information, forming a closed-loop cycle of perception, decision-making, actuation, and environmental response. Within this framework, neither information nor physical dynamics can be regarded as secondary to the other; instead, they co-evolve as mutually dependent components of a single nonequilibrium system.

External fields simultaneously provide energy and information. Information shapes interaction networks and collective-state transitions, while the resulting collective dynamics generate new correlations, structures, and memory states that subsequently influence future evolution. Through repeated iterations of this feedback process, active matter acquires the capacity to adapt, reorganize, and regulate its behavior across multiple scales. The resulting collective intelligence is therefore not solely a computational property nor purely a physical phenomenon, but an emergent consequence of the continuous coupling between information flow and nonequilibrium dynamics.

Viewed from this perspective, field-driven robotic active matter represents more than a collection of interacting robots. It constitutes a new class of programmable physical systems in which energy flow, information processing, and material dynamics become inseparably integrated. This unified framework provides the conceptual foundation for understanding how collective intelligence can ultimately be translated into functional material properties. Building upon this perspective, the next section explores how the interplay among fields, information, and collective dynamics can be systematically transformed into adaptive material functionalities and engineering applications.

\section{\label{sec:Material}From Fields to Information to Physics to Function: Materialization of Collective Intelligence}

Building upon the multifunctional roles of fields (Section \ref{sec:fields}), their physical realization in robotic active matter (Section \ref{sec:scales}), the emergence of programmable nonequilibrium phenomena (Section \ref{sec:nonequi}), and the unified physics–information coupling framework (Section \ref{sec:Coupling}), a central engineering challenge naturally arises: how can the collective physics of field-driven active matter be systematically transformed into useful material functionalities? Unlike conventional materials, whose properties are primarily determined by composition and structure, robotic active matter derives its behavior from the dynamic interplay among external fields, information processing, and collective dynamics. Consequently, engineering intelligent matter requires not only the design of constituent agents but also the design of mappings that translate control signals into functional material responses\cite{kaspar2021rise}. To address this challenge, we propose a general engineering pathway: "Field to Information to Dynamics to Function to Material Property". Within this framework, external and environmental fields provide both energetic and informational inputs; information-processing mechanisms convert these inputs into collective dynamical states; collective dynamics give rise to reusable functional modes; and these functional modes ultimately materialize as macroscopic material properties. This perspective provides a unified design principle for translating collective intelligence into engineering functionality.\\

\subsection{\label{subsec:FieToDyn}From Fields to Physics: Engineering Control Through Information-Mediated Collective States}

The first stage of the mapping concerns how externally applied and environmentally generated fields are converted into collective dynamical states. In practical applications, fields serve not only as sources of actuation but also as carriers of task-relevant information.

Programmable external fields—including magnetic, electric, acoustic, optical, and fluidic fields—provide globally accessible control variables through their amplitude, frequency, phase, spatial gradients, and temporal modulation\cite{yu2023programmable}. These fields define controllable interaction and information landscapes that coordinate large populations of agents while simultaneously supplying energy for collective motion. Through appropriate field design, collective systems can be driven into distinct dynamical regimes, including clustering, synchronization, coherent transport, phase separation, and dynamic jamming\cite{law2023micro}.

Real-world environments introduce an additional class of fields that function primarily as information sources rather than driving mechanisms. Temperature gradients, chemical gradients, pressure distributions, shear flows, obstacle-induced stress fields, and boundary-induced geometric constraints continuously provide information about environmental conditions. Robotic active matter equipped with sensing, memory, and feedback capabilities can convert these environmental signals into adaptive behavioral responses\cite{takatori2025feedback,palagi2018bioinspired}. Consequently, collective dynamics emerge not only from external control protocols but also from continuous interactions between the swarm and its environment.

Information serves as the critical intermediate layer connecting fields to dynamics. Local sensing, memory states, and feedback policies transform environmental and externally programmed signals into modifications of interaction rules and collective behavior\cite{wang2024harnessing}. In this way, information biases how a system explores its dynamical state space and selects collective states appropriate for specific tasks. From an engineering perspective, the primary objective is therefore not merely to actuate robotic matter, but to convert environmental information into desired collective dynamical modes through programmable field-mediated control. These collective dynamical modes provide the physical substrates from which functional primitives can emerge.

\begin{figure}[t]
    \centering
\includegraphics[width=1.0\linewidth]{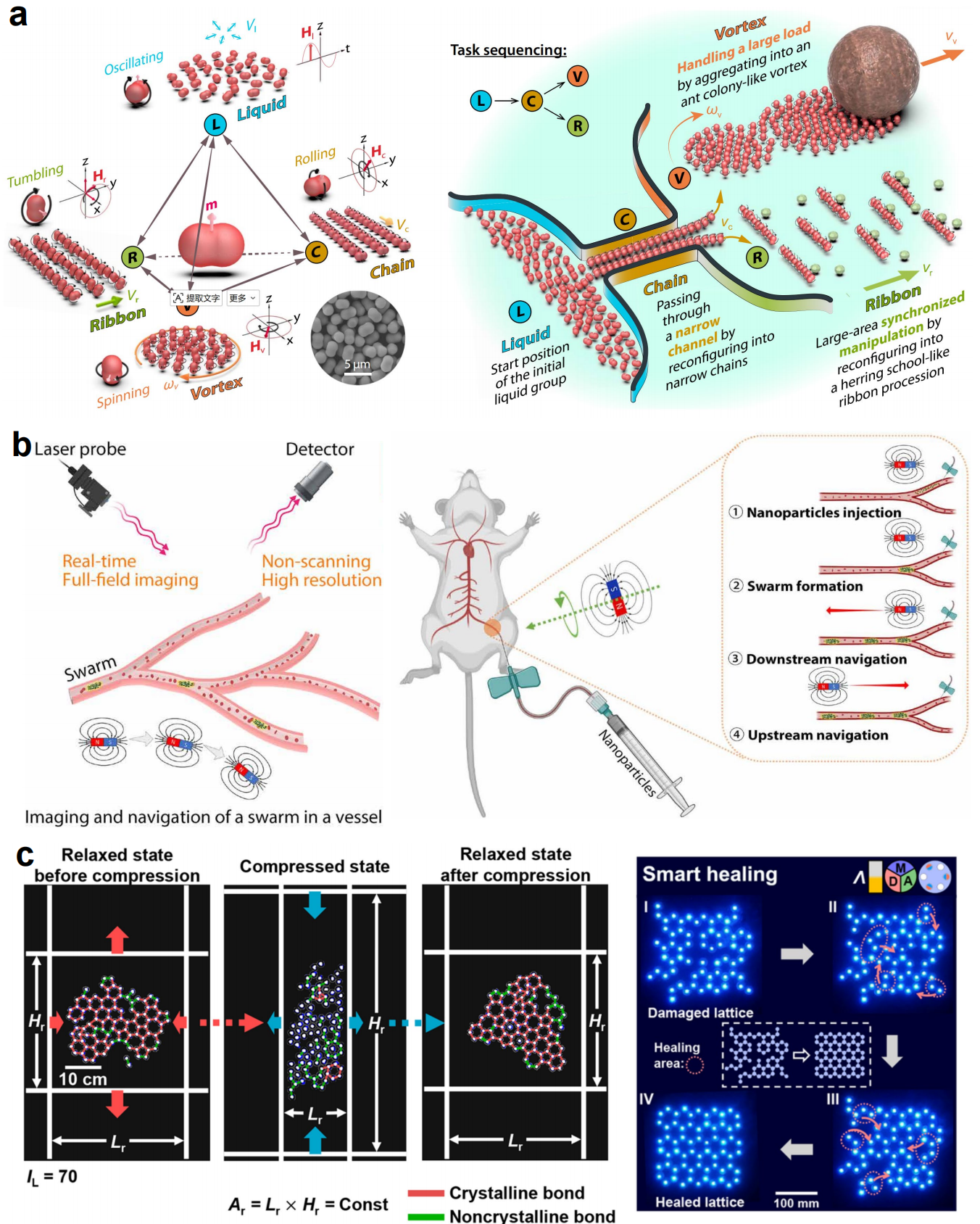}
\caption{\textbf{Material functions enabled by collective dynamics of field-driven robotic active matter.} (a) The reconfigurable hematite colloidal microrobots energized by alternating magnetic fields can transform between four programmable collective formations to accomplish multiple tasks\cite{xie2019reconfigurable}. Reproduced with permission from Xie \textit{et al.}, Sci. Robot. \textbf{4}(28), eaav8006 (2019). Copyright 2019 The American Association for the Advancement of Science. (b) Based on laser speckle contrast imaging, magnetic-field-driven microrobotic swarms can be real-time tracked and navigated in stagnant and flowing blood environments to achieve active targeted delivery \textit{in vivo}\cite{wang2024tracking}. Reproduced with permission from Wang \textit{et al.}, Sci. Robot. \textbf{9}(87), eadh1978 (2024). Copyright 2024 The American Association for the Advancement of Science. (c) Robo-Matter composed of light-field-driven Magbots can achieve spontaneous or controlled adaptive self-repair\cite{wang2024robo}. Reproduced from Wang \textit{et al.}, Nat. Commun. \textbf{15}(1), 8853 (2024); licensed under a Creative Commons Attribution License 4.0 (CC BY).}
\label{materials}
\end{figure}

\subsection{\label{subsec:DynToFun}From Physics to Function: Collective Dynamics as Functional Primitives}

Collective dynamics become engineering-relevant when they can be harnessed as experimentally demonstrated and reusable functional primitives. Rather than viewing clustering, synchronization, jamming, collective transport, or self-organization solely as physical phenomena, these nonequilibrium dynamical states can provide direct mechanisms for structural reconfiguration, mass transport, force generation, self-repair, and information processing. Importantly, these functionalities do not arise independently of collective dynamics; rather, they emerge from specific dynamical modes whose properties can be externally programmed through fields and system parameters.

\textbf{Mechanical functionality via collective force generation.} Collective dynamics can transform microscopic activity into macroscopic mechanical properties. At the material scale, gram-scale suspensions of light-triggered chemical nanomotors have been shown to exhibit an activity-dependent, reversible change in bulk viscosity, demonstrating that field-driven collective activity can directly regulate the rheological properties of an active medium\cite{choudhury2019chemical}. At the robotic scale, gradient magnetic fields can assemble microrobots into robust reconfigurable collectives with strongly enhanced force output, enabling manipulation of droplets, fluids, and solid objects\cite{xu2025reconfigurable}. These systems illustrate a feasible route from collective dynamics to function, in which the collective state amplifies microscopic actuation into macroscopic mechanical response.

\textbf{Reconfiguration via collective phase transitions.} Field-controlled transitions between distinct collective states, such as fluid-like, solid-like, and intermediate states, provide a general mechanism for structural reconfiguration. For example, as shown in Fig. \ref{materials}(a), the hematite colloidal microrobots energized by alternating magnetic fields can reversibly transform among liquid-, chain-, vortex-, and ribbon-like configurations, enabling collective morphology to be dynamically adapted to different tasks\cite{xie2019reconfigurable}. By dynamically reorganizing collective architectures, robotic matter can switch between shape adaptation, stiffness modulation, and morphology locking\cite{pishvar2020foundations,zhang2023magnetoactive,yan2012linking}. For example, millimeter-scale robotic collectives incorporating phase-change materials and magnetic particles can reversibly transition between rigid structural states and highly deformable fluidic states through photothermal or magnetic control\cite{wei2024magnetoactive,wang2023magnetoactive}. Similar field-controlled transitions have been demonstrated in synchronized active spinner materials, where collective rotation and synchronization produce reconfigurable structures with tunable transport properties\cite{han2020reconfigurable}. Programmable chiral states in magnetic roller flocks further demonstrate that collective states can be switched remotely by modulating the phase of the driving field, thereby converting dynamical reconfiguration directly into controllable pumping\cite{han2021programmable}.

\textbf{Transport via cooperative collective motion.} Collective modes such as vortices, coherent flows, and coordinated migration naturally provide direct routes from nonequilibrium dynamics to practical transport functionality. These dynamical states enable directional cargo delivery, large-area coverage, and targeted resource distribution\cite{gardi2022microrobot,yang2025magnetic,yang2026collective,zhao2026vivo}. Magnetically twisted colloidal ribbons can generate hydrodynamic flow fields capable of trapping, transporting, and releasing microscopic particles, demonstrating that propulsion of a collective structure can be directly converted into fluid-mediated cargo transport\cite{massana2017propulsion}. Colloidal swarms have likewise been shown to capture and transport passive cargos through emergent collective flows\cite{yang2020cargo}, while self-assembled magnetic colloidal swarms can dynamically organize into pump architectures with externally switchable flow direction\cite{han2024self}. More recently, confinement-induced unraveling of microrobot swarms has produced artificial microtubule-like structures that support directed cargo transport under confinement, including in flowing blood\cite{wang2026artificial}. For \textit{in vivo} applications, bio-inspired magnetic microrobot swarms can switch between compact vortex configurations optimized for long-range transport and extended ribbon-like structures that maximize infiltration and surface coverage within biological tissues\cite{xie2019reconfigurable,cai2025multifunctional,wang2021trends}. For example, as shown in Fig. \ref{materials}(b), based on laser speckle contrast imaging, magnetic-field-driven microrobotic swarms can be real-time tracked and navigated in stagnant and flowing blood environments to achieve active targeted delivery \textit{in vivo}\cite{wang2024tracking}. These examples illustrate a general principle: collective motion can act not merely as locomotion of the agents themselves, but as a functional transport field that reorganizes the surrounding environment and mediates cargo motion.

\textbf{Self-repair via adaptive reorganization.} Collective systems can also exploit defect migration, local reassembly, and feedback-driven restructuring to recover functionality after damage. Unlike conventional passive materials, repair processes can emerge from active collective responses, enabling distributed recovery and fault tolerance without centralized intervention\cite{brandenbourger2019non}. In other words, collective active dynamics can provide a fundamentally different route to self-repair, in which structural recovery results from the continual rearrangement of active constituents rather than from passive chemical or microstructural healing. Magnetically propelled colloidal carpets, for example, can undergo field-controlled treadmilling in which particles detach from the rear, migrate across the structure, and reattach at the advancing front, enabling rapid regeneration of damaged or disordered regions\cite{massana2019tunable}. Such dynamics demonstrate how continuous energy consumption and collective particle exchange can sustain structural integrity far from equilibrium. Related reconfigurable microswarms further show how collective assembly can restore functional architectures after structural disruption. For example, as shown in Fig. \ref{materials}(c), Robo-Matter composed of light-field-driven Magbots can achieve spontaneous or controlled adaptive self-repair without onboard computation or communication\cite{wang2024robo}. These examples suggest that self-repair in active matter can be understood as a dynamic maintenance process, rather than a one-time material healing event.

\textbf{Information processing through collective dynamics.} Beyond mechanical functions, collective dynamics can themselves serve as physical substrates for information processing. Synthetic active particles have recently been demonstrated as physical reservoir computers, in which self-organized nonlinear dynamics provide fading memory and high-dimensional state representations for predictive computation\cite{wang2024harnessing}. This example extends the function of collective active matter from manipulating matter in physical space to processing information in dynamical state space. In this sense, collective dynamics can simultaneously serve as mechanical functionality and computational functionality, providing a direct bridge between active matter, embodied computation, and intelligent materials.

Importantly, information processing enhances each of these functional primitives. Feedback improves transport accuracy, memory stabilizes desired configurations, and adaptive sensing enables context-dependent switching among functional modes\cite{takatori2025feedback}. Consequently, functionality emerges from the combined action of collective physics and information processing rather than from physical dynamics alone.

\subsection{\label{subsec:FunToMat}From Function to Material Properties: Materialization of Collective Intelligence}

The examples above reveal that material functions in robotic active matter need not be encoded statically in the composition or geometry of a material. Instead, they can emerge dynamically from collective states that are continuously generated, switched, and maintained by external fields and information-mediated interactions. Collective force generation produces tunable mechanical response; reconfiguration provides structural programmability; collective transport generates active pumping and cargo-handling capabilities; adaptive reorganization enables dynamic self-repair; and nonlinear collective dynamics can provide physical substrates for information processing. Thus, the functional primitives identified above constitute intermediate dynamical layers through which collective physics is materialized into effective material properties.

The final stage of the mapping transforms functional primitives into macroscopic material properties. The same functional primitive can materialize into multiple classes of engineering materials, while a given material property may arise from the combination of several functional primitives. At this level, robotic active matter becomes a new class of programmable material whose effective properties emerge from collective organization and information-guided dynamics. In these systems, material properties are no longer fixed consequences of composition alone. Instead, they emerge from dynamically programmable collective states that can be continuously reconfigured according to task requirements and environmental conditions.

\textbf{Adaptive mechanical materials.} By exploiting reconfiguration and dynamic mechanical reinforcement, robotic matter can continuously adjust its effective stiffness, compliance, information transport and load-bearing characteristics\cite{shankar2017topological}. Such materials are particularly attractive for soft robotic systems, minimally invasive surgical devices, and adaptive protective structures, where mechanical requirements vary during operation\cite{zheng2025metamaterial}.

\textbf{Reconfigurable architected materials.} Structural reconfiguration enables persistent shape adaptation and functional switching. Field-driven robotic active matter provides a route toward architected materials whose geometry can be dynamically programmed rather than statically prescribed. Such reconfigurable structures could enable morphing aerospace systems, adaptive architectural systems that regulate environmental exposure, and robotic metamaterials whose geometry could be actively modified in response to changing operating conditions\cite{mcevoy2015materials,melio2026pivoting}.

\textbf{Active transport materials.} Collective transport primitives can be materialized into intelligent transport media capable of capture, delivery, and controlled release. At microscopic scales, magnetic robotic swarms can perform targeted drug delivery, pollutant collection, and cargo manipulation\cite{landers2025clinically,te2025metareview,xu2025reconfigurable,yu2018ultra,wang2021ultrasound,ussia2024magnetic}. In biomedical applications, biohybrid robotic collectives have demonstrated substantially enhanced accumulation at target sites compared with passive transport mechanisms\cite{patino2025swarming,del2023ultrasound,du2026microswarm,hwang2019catalytic}.

\textbf{Self-healing materials.} Adaptive repair primitives provide a foundation for self-healing material systems in which structural integrity is maintained through active collective reorganization rather than passive material recovery. Experimental demonstrations of field-driven colloidal carpets show that collective activity can sustain structural integrity and restore ordered configurations after disruption, providing a physical realization of self-repair based on continuous energy consumption and collective reorganization\cite{massana2019tunable}. At the device level, reconfigurable microswarms inspired by ant bridges have further been used to restore broken electrical circuits, illustrating how collective structural reorganization can directly recover functionality after disruption\cite{jin2019mimicking}. More generally, these examples suggest a distinct paradigm of self-healing materials in which repair is not encoded solely in material composition, but emerges as a dynamically maintained collective property enabled by field-driven continuous sensing, reconfiguration and repair\cite{yu2023machine}.

The engineering significance of field-driven robotic active matter extends beyond the creation of individual functional materials. More fundamentally, it introduces a new design paradigm in which material behavior is generated through programmable mappings linking fields, information, collective dynamics, and functionality. In conventional materials science, engineering design often follows the pathway from structure to property. By contrast, field-driven robotic active matter introduces an additional dynamical layer in which material properties emerge through programmable collective states. The resulting pathway can therefore be expressed as “Field to Information to Dynamics to Function to Material Property”, where function serves as the intermediate dynamical mechanism linking collective physics to macroscopic material behavior. Within this framework, material properties become programmable outputs of collective information-processing dynamics rather than static consequences of material composition. Engineering intelligent matter therefore shifts from fabricating fixed structures toward designing controllable transformations between information and functionality. This perspective provides a direct pathway for translating collective intelligence into practical engineering systems and establishes the conceptual foundation for the quantitative benchmarking and performance evaluation framework developed in the following section.

\section{\label{sec:benchmarking}Benchmarking and Quantification}

To transition field-driven active matter from phenomenological laboratory demonstrations toward quantitatively engineered intelligent materials, a unified benchmarking framework is required. Such a framework must evaluate not only physical performance but also the ability of a system to transform information into collective dynamics, functional behaviors, and ultimately useful material properties. Building directly upon the hierarchical mapping established in Section \ref{sec:Material}—Field to Information to Dynamics to Function to Material Property—we propose that benchmarking should be organized according to the same engineering pathway. Rather than evaluating field-driven active matter solely through conventional metrics such as speed or force output, performance assessment should quantify how effectively collective intelligence is translated into functional and material capabilities. Within this perspective, benchmarking can be divided into four complementary layers: dynamical benchmarking, functional benchmarking, material-property benchmarking, and task-level benchmarking. Together, these layers provide a common performance reference system for intelligent robotic matter across different physical mechanisms, length scales, and application scenarios.

\begin{table*}[t]
\caption{\label{tab:quantitative_benchmark}
Quantitative benchmarks for representative field-driven active-matter platforms discussed in this review.}
\begin{ruledtabular}
\begin{tabular}{p{2.5cm} p{2.0cm} p{2.0cm} p{1.8cm} p{6.0cm}}
Representative platform & 
Main function & 
Benchmark metric & 
Reported value$^{1}$ & 
Experimental context
\\
\hline

Anatase TiO$_{2}$ colloidal nanomotors~\cite{choudhury2019chemical}
&
Mechanical functionality
&
Viscosity modulation
&
$\sim 10\times$
&
The bulk viscosity of a gram-scale active medium can be reversibly modulated by changing the activity of light-triggered nanomotors.
\\

Fe-Ni alloy microrobot collectives~\cite{xu2025reconfigurable}
&
Mechanical functionality
&
Force enhancement
&
$\sim 700\times$
&
Gradient magnetic fields generate robust reconfigurable microrobot collectives with strongly enhanced force output; a $0.2$-g collective produces Newton-level output forces.
\\

Hematite colloidal microrobots ~\cite{xie2019reconfigurable}
&
Reconfiguration
&
Reconfiguration timescale
&
$1.6$--$128$~s
&
Programmable transformations among liquid, chain, ribbon, and vortex states.
\\

Ferromagnetic Ni microparticles~\cite{han2021programmable}
&
Reconfiguration
&
Reconfiguration timescale
&
$\sim25$~ms
&
The chirality of a self-organized vortex can be reversibly switched by a $\pi$ phase shift of the driving magnetic field. At $f=40$~Hz, one field cycle corresponds to $\sim25$~ms.
\\

Hematite colloidal ribbons~\cite{massana2017propulsion}
&
Transport
&
Transport speed
&
$\sim 2.5~\mu$m~s$^{-1}$
&
Magnetically propelled colloidal ribbons generate hydrodynamic flows capable of trapping and transporting nonmagnetic cargo particles.
\\

Ferromagnetic Ni spheres~\cite{han2024self}
&
Transport
&
Transport speed
&
$\sim 10$~mm~s$^{-1}$
&
Self-assembled magnetic colloidal swarms form reconfigurable pump architectures that generate globally directed flow and transport glass cargo particles.
\\

Paramagnetic colloidal carpets~\cite{massana2019tunable}
&
Self-repair
&
Self-repair time
&
$\sim 2$~min
&
A field-controlled treadmilling process restores an initially disordered colloidal carpet to a highly ordered crystalline state.
\\

Ferromagnetic Ni spinners ~\cite{han2020reconfigurable}
&
Self-repair
&
Self-repair time
&
A few seconds
&
A locally fractured spinner lattice, intentionally disrupted by a 3-mm glass bead, spontaneously restores its ordered structure within a few seconds.
\\

Polymer microbead reservoir computer~\cite{wang2024harnessing}
&
Information processing
&
Prediction accuracy (normalized root-mean-square error)
&
$\sim 0.6$
&
Normalized root-mean-square error can be used to characterize the free-running prediction performance of a self-organized nonlinear physical reservoir computing system—composed of light-field-driven synthetic active microparticles—for chaotic time series.
\\

\end{tabular}
\end{ruledtabular}

\vspace{2pt}
\begin{minipage}{\textwidth}
\footnotesize
\noindent
\raggedright
$^{1}$ Only experimentally reported quantities are included; values are not directly compared across systems when the underlying definitions or operating conditions differ. Values are reported under the specific operating conditions of the corresponding studies and should therefore be interpreted as representative benchmarks rather than universal performance limits. Where a metric has no standardized definition across platforms, the experimentally reported quantity is retained without additional normalization.
\end{minipage}
\end{table*}

Importantly, several field-driven active-matter platforms already report quantitative performance measures that can be mapped onto these benchmarking layers. Table~\ref{tab:quantitative_benchmark} summarizes representative experimental values for mechanical functionality, reconfiguration, transport, self-repair, and information processing. These values demonstrate that the proposed framework can be grounded in experimentally measurable quantities, while also highlighting the current lack of standardized definitions and cross-platform testing protocols.\\

\subsection{\label{subsec:DynaBench}Dynamical Benchmarking: Quantifying Collective Physical Performance}

The first layer evaluates the collective dynamical states generated through field-mediated physics–information coupling. Since collective dynamics constitute the physical substrate from which higher-level functions emerge, their performance determines the accessible operational space of robotic matter.

For transport-oriented systems, dynamical performance is commonly characterized by transport speed $v$, often normalized by body length per second, together with transport efficiency and resistance to environmental disturbances\cite{zhou2021magnetically}. In complex physiological fluids or unsteady shear environments, effective locomotion is constrained by the mechanical power required to overcome external resistance. Representative magnetic microrobotic systems have already achieved high-speed propulsion together with upstream navigation in physiologically relevant flow environments\cite{yang2022functionalized,mohanty2022ceflowbot}.

For reconfigurable collectives, reconfiguration timescale and response latency provide direct measures of dynamical agility. High-performance acoustically controlled systems are capable of undergoing complex structural reconfiguration on millisecond timescales, enabling rapid adaptation to dynamically changing environments\cite{xiao2022acoustics,zhang2024sonotransformers}. Metrics such as synchronization quality, collective coherence, clustering efficiency, and dynamic stability further characterize the robustness of collective-state formation.

Because information participates directly in collective-state selection, benchmarking at this level must also incorporate informational descriptors. Correlation length $\xi$, mutual information, information flow, entropy-production rate, and response functions provide quantitative measures of information propagation, collective sensitivity, and self-organization capability\cite{shaebani2020computational,ro2022model}. Near critical regimes, correlation lengths may become comparable to the system size, enabling efficient transmission of weak signals and enhanced responsiveness to environmental inputs\cite{cavagna2010scale}. Collectively, these metrics define the dynamical workspace within which robotic matter can operate.

\subsection{\label{subsec:FunBench}Functional Benchmarking: Quantifying Information-Enabled Functional Primitives}

While dynamical metrics characterize physical behavior, engineering applications ultimately depend on functional performance. The second benchmarking layer therefore evaluates the functional primitives introduced in Section \ref{sec:Material} and quantifies how effectively collective dynamics are converted into useful functionality.

For reconfiguration functionality, benchmarking focuses on shape-change range, reconfiguration speed, reversibility, and structural stability\cite{yoder2024hexagonal,cheng2025magnetic,dai2024research}. These metrics quantify the ability of robotic matter to transition between distinct structural states while maintaining functional integrity.

For transport functionality, relevant measures include navigation success rate, transport throughput, target-capture efficiency, delivery precision, and environmental coverage efficiency\cite{bayindir2016review}. Such metrics directly characterize the effectiveness of cooperative motion in accomplishing transport tasks.

For self-repair functionality, key indicators include recovery ratio, repair time, fault tolerance, and functionality restoration following damage\cite{peck2022self}. Unlike conventional passive materials, repair processes emerge from active collective responses; therefore both recovery speed and recovery quality are critical performance measures.

Information processing plays a central role in shaping functional performance. Navigation accuracy depends on sensing quality and information propagation, adaptation speed reflects the efficiency of feedback control, and robustness under uncertainty is closely related to memory and decision-making capability\cite{refis2025network}. Consequently, functional benchmarking evaluates not only physical execution but also the effectiveness with which information is transformed into task-oriented behavior.

\subsection{\label{subsec:MaterBench}Material-Property Benchmarking: Quantifying Emergent Intelligent Material Behavior}

The third layer evaluates the macroscopic material properties that emerge from collective functionality. At this level, field-driven robotic active matter is assessed as a material system rather than as a collection of individual robots.

For adaptive mechanical materials, key metrics include effective modulus $G$, soft-to-rigid switching ratio, load-bearing capacity, damping performance, viscosity modulation, force enhancement, and mechanical stability under repeated operation\cite{pishvar2020foundations}. Achieving stiffness modulation spanning several orders of magnitude remains a major objective for programmable robotic materials\cite{zhang2023bioinspired}.

For reconfigurable architected materials, relevant measures include shape programmability, structural persistence, environmental adaptability, and geometric complexity\cite{thalamy2019survey,jin2021collective}. These metrics quantify the extent to which material architectures can be actively reconfigured while retaining functionality.

For active transport materials, benchmarking focuses on transport capacity, delivery precision, release controllability, and transport efficiency under environmental constraints\cite{nedjah2019review}. Such measures characterize the effectiveness of robotic matter as an intelligent transport medium.

For self-healing materials, evaluation emphasizes healing efficiency, conductivity recovery ratio, repeatability, and long-term durability\cite{bartlett2019self,heinrich2019swarm}. For example, liquid-metal composite systems designed for self-healing applications may be required to maintain near-lossless electrical conductivity recovery after dozens of damage–repair cycles\cite{markvicka2018autonomously}.

An important distinguishing feature of intelligent materials is that material performance depends not only on mechanical efficiency but also on information utilization. Consequently, coupled metrics such as information-to-work efficiency, information-to-function efficiency and prediction accuracy become increasingly important. These quantities characterize how effectively sensing, memory, and decision-making processes are converted into useful mechanical output and material functionality. From the perspective of information thermodynamics, the Landauer limit, $k_B T \ln 2 \approx 2.9 \times 10^{-21}$ J/bit at room temperature, provides a theoretical lower bound on the energetic cost of irreversible information processing\cite{landauer1961irreversibility,bennett1982thermodynamics}. Although practical robotic systems operate far above this limit, it establishes a universal reference for evaluating the efficiency of physics–information integration in intelligent matter.

\subsection{\label{subsec:StandBench}Task-Level and Cross-Platform Benchmarking}

Ultimately, benchmarking must enable meaningful comparisons across diverse robotic-matter platforms and facilitate translation into real-world engineering applications. The fourth benchmarking layer therefore focuses on task-level performance and cross-platform evaluation.

Standardized task suites provide hardware-independent benchmarks that evaluate the practical capabilities of robotic matter under controlled conditions. Representative examples include maze navigation, distributed cargo sorting, adaptive reconfiguration, collective assembly, and self-repair tasks\cite{lian2021benchmarking}. Such benchmarks allow systematic comparison of robustness, adaptability, and scalability across different physical implementations.

Existing industrial robotics benchmarking frameworks, such as RobotPerf$^{\rm TM}$ software benchmarking suite and physical benchmarking frameworks developed by NIST, offer useful references for constructing standardized evaluation protocols\cite{mayoral2024robotperf,bi2019overview}. In safety-critical domains including aerospace systems, implantable medical devices, and autonomous infrastructure, benchmarking must additionally incorporate long-term repeatability testing, reliability analysis, and failure mode and effects analysis.

Future intelligent materials may require entirely new performance descriptors that have no direct counterparts in conventional materials science. Examples include a morphological adaptability factor, quantifying the ability of a material to maintain functionality under environmental perturbations such as temperature variations, channel blockage, or mechanical damage, and a generalized task-capability index, measuring the number of heterogeneous tasks executable by a single physical platform through modifications of information protocols alone, without hardware redesign\cite{nakajima2015information}.

Through this hierarchical benchmarking framework, field-driven robotic active matter can be evaluated not merely as active materials or robotic systems, but as quantitatively engineered information-processing matter. Such a framework provides the necessary foundation for comparing diverse intelligent-material platforms, accelerating technological deployment, and establishing the performance standards required for the next generation of autonomous functional materials.\\

\section{\label{sec:challenges}Challenges and Future Directions}

Despite rapid advances in field-driven active matter, significant challenges remain before such systems can evolve from laboratory-scale demonstrations into practical intelligent materials. The preceding sections have established a hierarchical pathway linking fields, information, collective dynamics, functionality, and material properties, together with a quantitative benchmarking framework for evaluating system performance. The next frontier is therefore not merely improving individual demonstrations, but realizing robotic matter that is simultaneously scalable, adaptive, and theoretically understood. Achieving this objective requires progress along three interconnected directions: the development of scalable intelligent matter, the emergence of adaptive autonomous matter, and the establishment of a fundamental theory of information-driven nonequilibrium systems.\\

\subsection{\label{subsec:ScalableMatter}From Programmable Control to Scalable Intelligent Matter}

One of the most immediate engineering challenges is scalability. While many existing robotic active-matter platforms successfully demonstrate collective behaviors using tens to hundreds of agents, practical intelligent materials may ultimately require populations ranging from thousands to millions of interacting units. Maintaining precise collective control while preserving scalability represents a major technological bottleneck\cite{miskin2020electronically,lassiter2025microscopic}.

A central difficulty arises from the increasing complexity of field-mediated interactions. As system size grows, collective behavior becomes progressively more sensitive to fluctuations, communication delays, hydrodynamic coupling, and environmental heterogeneity. The incorporation of multiple control modalities—including acoustomagnetic, optoelectrical, and magnetohydrodynamic actuation—provides expanded control degrees of freedom but simultaneously introduces highly nonlinear coupling among mechanical, fluidic, and informational processes\cite{wang2025advanced,roy2019effect}. Developing scalable control architectures capable of coordinating large populations without sacrificing robustness therefore remains a critical engineering objective.

Recent advances in artificial intelligence (AI) provide promising approaches for addressing the scalability challenge by enabling decentralized and adaptive control of large-scale active-matter systems. Conventional control strategies for robotic collectives often rely on manually designed interaction rules or centralized optimization, whose complexity increases rapidly with the number of agents. In contrast, reinforcement learning and multi-agent learning approaches can allow individual agents to infer effective behavioral policies from local sensing and environmental feedback, thereby reducing the need for explicit global coordination\cite{blais2023reinforcement,heuthe2024counterfactual}. Such learning-based control architectures may enable large populations of field-driven agents to autonomously adapt their interaction rules, compensate for environmental fluctuations, and maintain collective functionality despite communication limitations and uncertainty. Beyond optimizing predefined collective states, AI may also provide a pathway toward discovering new collective behaviors that are difficult to design analytically\cite{cai2025reinforcement,liu2026efficient}. By coupling learning algorithms with physical field-mediated interactions, future robotic active matter may achieve scalable self-organization in which collective structures and functions emerge through continuous interaction between agents, fields, and environments. However, integrating AI into large-scale active matter introduces additional challenges, including computational cost, energy consumption, interpretability of learned control strategies, and compatibility between digital learning processes and physical nonequilibrium dynamics.

Energy constraints introduce an additional limitation, particularly at micro- and nanoscale dimensions. Sensing, state updating, communication, and information processing all consume finite energetic resources\cite{zhou2020enhanced}. According to Landauer's principle, the irreversible erasure of one bit of information requires a minimum energy dissipation of approximately $k_B T \ln 2 \approx 2.9 \times 10^{-21}$ J at room temperature\cite{landauer1961irreversibility}. Although real active systems operate far above this limit, information processing nevertheless contributes to the overall energetic budget of the material. Consequently, future robotic matter must balance sensing accuracy, control complexity, and energy efficiency within finite physical constraints. Understanding these tradeoffs will be essential for constructing large-scale intelligent materials capable of long-term operation in realistic environments.

\subsection{\label{subsec:AdaptiveMatter}From Reactive Matter to Adaptive Intelligent Matter}

Beyond scalability, future robotic matter must become increasingly autonomous. Most existing field-driven systems rely on externally predefined control protocols that specify collective behavior in advance. While such approaches enable precise actuation, they remain fundamentally reactive and often struggle in unpredictable environments.

A major future direction is the integration of local field programmability with machine learning, adaptive control, and distributed decision-making algorithms. By combining field-mediated interactions with reinforcement learning, distributed policy optimization, and real-time feedback, robotic matter may acquire the ability to continuously update its behavior in response to environmental changes\cite{liu2025embodied,muinos2021reinforcement}. Within such architectures, collective dynamics would no longer be solely determined by externally imposed control signals, but would emerge through continuous interactions among sensing, learning, and physical embodiment.

Recent advances in AI, particularly foundation models and embodied AI, provide new opportunities to further extend adaptive robotic matter beyond task-specific learning. While reinforcement learning and distributed optimization can enable agents to adapt their behaviors within predefined objectives, large-scale multimodal models offer the possibility of integrating heterogeneous sensory information, prior knowledge, and environmental context into collective decision-making processes\cite{brohan2023can,brohan2023rt}. By coupling such AI architectures with field-mediated physical interactions, future robotic matter may acquire the ability not only to optimize predefined behaviors but also to interpret complex environments, formulate adaptive strategies, and continuously improve its collective responses through experience. Importantly, the integration of AI into robotic active matter should not be viewed as replacing physical intelligence with external computation. Instead, AI may provide a higher-level information-processing layer that interacts with embodied physical dynamics\cite{bright2026autoscidact}. Local sensing, field-mediated communication, and collective morphology can serve as distributed computational substrates, while AI algorithms provide mechanisms for learning, prediction, and adaptive regulation. Such integration could establish a new paradigm of embodied intelligent matter, in which intelligence emerges from the closed-loop interaction among artificial cognition, physical fields, and collective dynamics.

This transition naturally connects to the broader concepts of embodied intelligence and morphological computation discussed in earlier sections. Future intelligent materials may actively exploit structural deformation, defect migration, and collective reconfiguration as computational resources\cite{paul2006morphological}. Materials could compensate for sensing uncertainty through adaptive morphology, reorganize collective structures to maintain functionality under perturbations, and autonomously trigger self-healing processes following damage\cite{zheng2025metamaterial,cui2022design,rus2015design}. In this sense, information processing would become inseparable from material behavior itself.

However, realizing adaptive intelligent matter introduces new challenges. In distributed systems lacking centralized control, maintaining collective coherence while preserving local adaptability remains difficult. Excessive adaptability may destabilize collective organization, whereas excessive coordination may suppress learning and responsiveness. Developing mechanisms that simultaneously support robustness, adaptability, and long-range consensus therefore represents one of the key challenges for next-generation robotic matter\cite{brambilla2013swarm}. Solving this problem could ultimately enable autonomous materials capable of self-guided navigation through unknown environments, adaptive operation in complex fluidic networks, and long-term functionality in changing conditions without continuous human intervention.

\subsection{\label{subsec:TheoryMatter}Toward a Theory of Information-Driven Nonequilibrium Matter}

At a deeper scientific level, many fundamental questions concerning the relationship between information and nonequilibrium physics remain unresolved. Throughout this review, information has been treated not merely as a control signal but as an active dynamical variable capable of influencing collective behavior. Establishing a rigorous theoretical framework for this perspective represents a major long-term challenge.

A central objective is the development of a quantitative bridge connecting information theory, stochastic thermodynamics, and nonequilibrium statistical mechanics\cite{fodor2022irreversibility,o2022time}. Such a framework would enable principled descriptions of how information is generated, propagated, stored, and converted into physical work within many-body active systems. One particularly important open question concerns the relationship between microscopic information flow and macroscopic collective behavior. While correlations, memory, and feedback clearly influence collective dynamics, a general theory linking information-processing measures to experimentally observable fluctuations and phase transitions remains largely absent.

Another fundamental challenge concerns the emergence of new universality classes in information-coupled active matter. Classical nonequilibrium systems are often classified according to conservation laws, symmetries, and hydrodynamic modes. In field-driven robotic matter, however, additional ingredients—including memory, feedback, adaptive decision-making, and programmable interactions—may fundamentally alter collective critical behavior\cite{shaebani2020computational}. Whether information-driven phase transitions constitute genuinely new universality classes remains an open question.

Related issues arise in systems exhibiting nonreciprocal interactions and unconventional mechanical responses. Phenomena such as odd elasticity, odd viscosity, active synchronization, and programmable criticality have already demonstrated behaviors that cannot be fully described within traditional equilibrium frameworks\cite{o2022time}. Understanding the phase boundaries, critical fluctuations, and self-organized critical mechanisms of these emerging states may reveal entirely new forms of nonequilibrium organization.

Ultimately, the long-term scientific vision extends beyond understanding active matter itself. A fundamental question is whether information can be incorporated as an effective thermodynamic state variable of nonequilibrium matter, analogous to energy, momentum, or particle density\cite{takatori2015towards}. Answering this question would establish a universal theoretical foundation for information-driven active systems and provide a unified language for describing both living collectives and artificial robotic matter. Such a framework could ultimately transform information from a descriptive quantity into a fundamental physical principle governing the behavior of intelligent matter.

\section{\label{sec:conclusion}Conclusion and Outlook: Toward Information-Driven Intelligent Active Matter}

In this Review, we have argued that the evolution of active matter is increasingly characterized by the integration of energy, information, and dynamics. Beginning with the multifunctional roles of external fields as sources of energy injection, interaction mediation, and information encoding, we examined how robotic swarms across multiple length scales provide physical realizations of field-mediated physics–information coupling. We further discussed how programmable nonequilibrium phenomena emerge from the interplay among fields, collective dynamics, and information processing, and proposed a unified framework linking information and physics through bidirectional feedback. Building upon this perspective, we established a hierarchical engineering pathway—“Field to Information to Dynamics to Function to Material Property”—that connects microscopic collective behavior to macroscopic material functionality, together with a benchmarking framework for quantitatively evaluating intelligent robotic matter. Collectively, these developments suggest that information should no longer be viewed merely as an external control signal, but rather as an active physical resource that participates directly in the organization and functionality of nonequilibrium matter.

Viewed from a broader perspective, the development of intelligent matter may be understood as part of a larger historical progression in the way functionality is embedded into physical systems. Traditional passive materials derive their properties primarily from composition and structure. Active matter introduced continuous energy consumption as a new mechanism for generating emergent behavior and adaptive responses. Field-driven active matter further expanded this paradigm by enabling the programmable regulation of collective dynamics through external fields\cite{yu2023programmable}. Looking forward, a new stage may be emerging: information-driven intelligent matter\cite{kaspar2021rise}. In this paradigm, information becomes a co-evolving dynamical quantity coupled to energy and matter. Material behavior is no longer determined solely by structure or energy dissipation, but increasingly by the generation, transmission, storage, and processing of information within collective systems\cite{miskin2026robotic}. The resulting materials are not merely responsive but increasingly capable of perception, adaptation, decision-making, and autonomous reconfiguration. In this sense, field-driven active matter represents not simply an extension of active matter physics, but a potential bridge toward a broader class of intelligent physical systems in which information and dynamics are intrinsically intertwined.

The ultimate scientific challenge is to establish a universal framework that places information on equal footing with traditional physical variables in nonequilibrium matter. As discussed in the previous section, an important open question is whether information can be treated as a thermodynamic state variable of active systems, analogous to energy, momentum, or particle density\cite{crosato2019irreversibility}. If such a framework can be achieved, future materials may be designed not only through composition, architecture, and energy flow, but also through programmable information flows that regulate collective behavior across scales. In this vision, sensing, memory, computation, learning, and adaptation would no longer be regarded as auxiliary functionalities added to materials, but as intrinsic material properties emerging from information–physics coupling. The long-term objective is therefore not merely to fabricate materials that perform predefined functions, but to engineer matter capable of perceiving, computing, adapting, and autonomously generating functions in response to changing environments. Realizing this vision would fundamentally transform materials science from the design of static structures into the engineering of intelligent matter and may ultimately provide a unified framework for understanding both artificial robotic collectives and living adaptive systems.\\

\begin{acknowledgments}
This work was supported by the National Natural Science Foundation of China (Grant Nos. T2350007, 12504235, and 12404239), seed grants from the Wenzhou Institute (WIUCASQD2021002).\\
\end{acknowledgments}

\section*{Author Declarations}

\textbf{Conflict of Interest:} The authors have no conflicts to disclose.\\

\textbf{Author Contributions:} Jing Wang, Shengkai Li, and Trung V. Phan contributed equally to this work.\\

\textbf{Jing Wang:} Conceptualization (equal); Formal analysis (equal); Investigation (equal); Visualization (equal); Writing – original draft (equal). \textbf{Shengkai Li:} Formal analysis (equal); Investigation (equal); Visualization (equal); Writing – original draft (equal). \textbf{Trung V. Phan:} Formal analysis (equal); Investigation (equal); Visualization (equal); Writing – original draft (equal). \textbf{Gao Wang:} Data curation (equal); Validation (equal); Writing – review \& editing (equal). \textbf{Huaicheng Chen:} Data curation (equal); Validation (equal); Writing – review \& editing (equal). \textbf{Xiaorui Chen:} Validation (equal); Writing – review \& editing (equal). \textbf{Robert H. Austin:} Supervision (equal); Writing – review \& editing (equal). \textbf{Liyu Liu:} Conceptualization (equal); Funding acquisition (equal); Supervision (equal); Writing – review \& editing (equal).\\

\textbf{Data Availability:} The data that support the findings of this study are available from the corresponding author upon reasonable request.

\bibliography{main}
\bibliographystyle{apsrev4-2}

\end{document}